\documentclass[a4paper,twosides]{article}
\usepackage{CJK,multicol,multirow,graphicx,fancyhdr,epstopdf}
\usepackage{amsmath,amsfonts,amssymb,bm,upgreek,mathrsfs,ccmap,mathcomp}
\usepackage[pagewise,switch,columnwise]{lineno}
\usepackage[compress,nospace]{cite}
\usepackage[dvipsnames]{xcolor}
\usepackage{CPL-2022}
\usepackage{threeparttable}

\newcommand{\cplyear}{2024} \newcommand{\cplvol}{39}
\newcommand{\cplno}{x} \newcommand{\cplpagenumber}{xxxxxx}
 \newcommand{\cplpage}{\cplpagenumber-\thepage}

\begin{document}

\begin{CJK}{GBK}{song}\vspace* {-4mm} \begin{center}
\large\bf{\boldmath{Statistical properties and cosmological applications of fast radio bursts}}
\footnotetext{\hspace*{-5.4mm}$^{*}$Corresponding author. Email: fayinwang@nju.edu.cn

\noindent\copyright\,{\cplyear}
\href{http://www.cps-net.org.cn}{Chinese Physical Society} and
\href{http://www.iop.org}{IOP Publishing Ltd}}
\\[5mm]
\normalsize \rm{}Qin Wu$^{1}$, Fa-Yin Wang$^{1,2*}$
\\[3mm]\small\sl $^{1}$School of Astronomy and Space Science, Nanjing University, Nanjing 210093, China

$^{2}$Key Laboratory of Modern Astronomy and Astrophysics, Nanjing University, Nanjing 210093, China 

\leavevmode \\[4mm]\normalsize\rm{}(Received xxx; accepted manuscript online xxx)
\end{center}
\end{CJK}
\vskip 1.5mm

\small{\narrower 
\par}\vskip 3mm
\normalsize\noindent{\narrower{PACS: 05.20.-y, 95.85.Bh, 98.70.Dk, 98.80.Es}}\\
\noindent{\narrower{DOI: \href{http://dx.doi.org/10.1088/0256-307X/\cplvol/\cplno/\cplpagenumber}{10.1088/0256-307X/\cplvol/\cplno/\cplpagenumber}}

\par}\vskip 5mm

\begin{abstract}
    Fast radio burst (FRB) is a type of extragalactic radio signal characterized by millisecond duration, extremely high brightness temperature, and large dispersion measure. It remains a mystery in the universe. Advancements in instrumentation have led to the discovery of 816 FRB sources and 7622 bursts from 67 repeating FRBs\footnote{https://blinkverse.alkaidos.cn/}. This field is undergoing rapid development, rapidly advancing our understanding of the physics of FRBs as new observational data accumulates. The accumulation of data has also promoted our exploration of our universe. In this review, we summarize the statistical analysis and cosmological applications using large samples of FRBs, including the energy functions, the waiting time distributions of repeating FRBs, probe of ``missing baryons" and circumgalactic medium in the universe, measurements of cosmological parameters, exploration of the epoch of reionization history, and study of the gravitational lensing of FRBs. 
\end{abstract}

\section{Introduction}

Since the first fast radio burst (FRB) was discovered in the archival data from the Parkes telescope in 2007 \cite{Lorimer2007}, the field of FRBs has been rapidly developed. The first-discovered FRB (also called Lorimer burst) had a peak flux density $S_{\nu}\gtrsim 30$\,Jy, a pulse duration $W\sim 5$\,ms and a dispersion measure ${\rm DM}\sim375\,{\rm pc\, cm^{-3}}$.  
In the years following the discovery of the Lorimer burst, no similar events were detected. Some suspected events were eventually confirmed as interference, known as perytons \cite{Burke-Spolaor2011}. 
In 2012, Keane et al\cite{Keane2012} reported the second highly dispersed burst with ${\rm DM}\sim946\,{\rm pc\, cm^{-3}}$. Thornton et al\cite{Thornton2013} discovered four FRBs from the Parkes telescope in 2013, confirming the FRB population's existence. 
Subsequently, other telescopes, including the 305\,m Arecibo telescope, first detected FRBs in 2014 \cite{Spitler2014}. The polarization properties were detected in 2015 \cite{Petroff2015,Masui2015}. 

Before 2016, all observed FRBs were non-repeating, until the first repeating bursts of FRB 20121102A were discovered \cite{Spitler2016}, confirming the existence of repeating FRBs. After that, the Very Large Array (VLA) and Arecibo identified the host galaxy of FRB 20121102A \cite{Chatterjee2017}. A compact persistent radio source was determined to be associated with the burst source using the European Very Long Baseline Interferometry (VLBI) Network and Arecibo \cite{Marcote2017}. Tendulkar et al\cite{Tendulkar2017} confirmed that FRB 20121102A is located in a dwarf-star-forming galaxy at redshift $z=0.19$. 

Soon after, radio telescopes such as the Canadian Hydrogen Intensity Mapping Experiment (CHIME) \cite{CHIME/FRB Collaboration2019a,CHIME/FRB Collaboration2019b}, the Australian Square Kilometre Array Pathfinder (ASKAP) \cite{KumarP2019}, and the Five-hundred-meter Aperture Spherical radio Telescope (FAST) \cite{Luo2020a,NiuCH2022} continuous monitored FRBs, leading to the discovery of an increasing number of repeating FRBs. 
Meanwhile, the localization of FRBs also experienced rapid advancement, with more and more host galaxies and redshifts of FRBs being confirmed using interferometric techniques by the ASKAP, Deep Synoptic Array (DSA), and other telescopes \cite{Bannister2019,Prochaska2019a,Ravi2019,Macquart2020,Marcote2020,Bhandari2022}. 
The increasing number of localized FRBs is significant for confirming the ${\rm DM_{IGM}}-z$ relationship \cite{Macquart2020}. 
Enhanced instrumentation significantly increased the quantity of detected FRBs and those with identified host galaxies. Continued advancements in technology facilitated the localization of numerous non-repeating FRBs. CHIME/FRB group released the first CHIME/FRB fast radio burst catalog, consisting of 536 FRBs \cite{CHIME/FRB Collaboration2021}.  

The CHIME/FRB Collaboration reported observations of a $16.35\pm 0.15$\,day periodicity from the repeating FRB 20180916B, with a five-day phase window \cite{CHIME/FRB Collaboration2020a}. Follow-up observations indicate that the bursts detected at higher frequencies tend to appear at slightly earlier phases than those detected at lower frequencies \cite{Pastor-Marazuela2021, Pleunis2021a}. A possible $\sim 157$\,day period was reported using 5\,yr data \cite{Rajwade2020}.
In addition to the long periods, CHIME/FRB Collaboration has found a short period of 216.8\,ms with a significance of 6.5$\sigma$ for FRB 20191221A \cite{CHIME/FRB Collaboration2022}. 

The origin of FRBs has been a mystery and most FRBs occur outside the Milky Way. 
FRB-like bursts FRB 20200428 from a galactic magnetar SGR J1935+2154 were detected by CHIME and the Survey for Transient Astronomical Radio Emission 2 (STARE2) on April 28th, 2020 \cite{CHIME/FRB Collaboration2020b,Bochenek2020}, which suggests that some FRBs may originate from magnetars. Meanwhile, these FRB-like bursts are associated with a hard X-ray burst \cite{LiCK2021,Mereghetti2020,Ridnaia2021,Tavani2021}. 
Subsequent long-term monitoring did not find similar events again, indicating that the coincidence of FRBs and X-ray bursts is rare \cite{Lin2020}. 
Based on the association of FRB 20200428 and magnetar SGR J1935+2154, the magnetar model becomes the leading source model \cite{Popov2010,Kulkarni2014,Katz2016,Lyubarsky2014,Metzger2017,Beloborodov2017,Beloborodov2020,Margalit2018,Metzger2019,Kumar2017,Yang2018,Wadiasingh2019,Wadiasingh2020,Lyubarsky2020,Zhang2022,Lyutikov2021,Thompson2023,Lu2020,Yang2021,Yang2020,WangFY2020,Zhong2020,Kremer2021}. 
Meanwhile, other models involving neutron stars have also been proposed to understand the radiation mechanism and origin of FRBs, including interacting neutron star models \cite{Katz2017}, comet/asteroid interaction models \cite{Geng2015,Dai2016,Bagchi2017,Smallwood2019,Dai2020,Dai2020b}, cosmic comb model \cite{Zhang2017}, binary comb models \cite{Ioka2020,Wada2021}, magnetospheric interaction models \cite{Hansen2001,Lai2012,Piro2012,WangJS2016}, and white-dwarf-fed neutron star model \cite{Gu2016,Gu2020}. Platts et al\cite{Platts2019} provided a living catalog for FRBs (the FRB theory Wiki page\footnote{https://frbtheorycat.org}), listing the majority of the models. A comprehensive analysis of various models was carried out in Zhang (2023)\cite{ZhangB2023}.

Recent advances in the field of FRBs have come from discoveries in the polarization properties. Diverse polarization angle swings are found from repeating FRB 20180301 \cite{Luo2020a}. A pulsar-like swing in the polarization angle of non-repeating FRB 20221022A was first detected by CHIME \cite{Mckinven2024}. 
Faraday rotation measure (RM) variations are found in most repeating FRBs \cite{Mckinven2023}, such as FRB 20201124A, FRB 20190520B and FRB 20180301A \cite{Xu2022,Anna-Thomas2023,Kumar2023}. This indicates the binary origin of these FRBs \cite{WangFY2022,Zhao2023,Rajwade2023}. 

There are several reviews summarizing the development of the field of fast radio bursts from different perspectives \cite{Katz2018,Popov2018,Katz2018a,Petroff2019,Cordes2019,ZhangB2020,Chatterjee2020,Caleb2021,Bhandari2021,Xiao2021,Bailes2022,ZhangB2023}. Xiao et al\cite{Xiao2021} summarized the basic physics of FRBs and discussed the research progress before 2021. Zhang\cite{ZhangB2023} gave a comprehensive review of the phenomenology and possible underlying physics of FRBs. 
Fast radio bursts have entered a rapid development period. An increasing number of FRBs are being discovered due to the large field of view of the CHIME telescope. The follow-up observations of repeating FRBs have accumulated a wealth of burst data using high-sensitivity telescopes, such as Arecibo and FAST. And the number of localized FRBs has significantly increased in recent years. 

More data drive the development of statistical analysis, leading to an increasing number of outstanding statistical works published and deepening our understanding of the origin of FRBs. On the other hand, the cosmological applications of FRBs are of significant importance for studying the composition of the universe, and advances in observations further propel theoretical work. These factors emphasize the urgent need to summarize the current progress in statistical analysis and cosmological applications in the field of FRBs. We will primarily summarize the current research on the statistical and cosmological applications of FRBs in the following sections.

\section{Intrinsic characteristics}
\subsection{Pulse width}
The typical pulse width of FRBs is on the order of milliseconds. In observations, the pulse width is typically defined as the width at 50\% or 10\% of the peak pulse intensity \cite{Lorimer2012}. 
The observed pulse width $W_{\rm obs}$ is composed of the intrinsic pulse width $W_{\rm int}$, scattering broadening during propagation $\tau_{\rm sc}$, and instrumental broadening $t_{\rm ins}$. The observed pulse width $W_{\rm obs}$ is \cite{ZhangB2023,Cordes2003,Lorimer2012}, 
\begin{equation}
    W_{\rm obs} = \sqrt{W_{\rm int}^2(1+z)^2+\tau_{\rm sc}^2+t_{\rm ins}^2},
\end{equation}
where $z$ is the redshift of FRB. The instrumental broadening is
\begin{equation}
    t_{\rm ins} = \sqrt{t_{\rm samp}^2+\Delta t_{\rm DM}^2+ \Delta t_{\rm DM_{err}}^2},
\end{equation}
where $t_{\rm samp}$ is the data sampling time, $\Delta t_{\rm DM}$ is the smearing due to dispersion measure
\begin{equation}
    \Delta t_{\rm DM} = 8.3\mu s{\rm DM}\Delta\nu_{\rm MHz}\nu_{\rm GHz}^{-3},
\end{equation} 
and $\Delta t_{\rm DM_{err}}$ is the smearing due to the error of dispersion measure \cite{Petroff2019,ZhangB2023}. 

Irregularities in the electron density or scattering properties of the medium cause a broadening of the temporal of the radio pulse during its propagation. The pulse profile of FRB generally exhibits a one-side exponentially decreasing tail, which is caused by the plasma scattering effect. This scattering tail is dependent on the frequency $\tau_{\rm sc}\propto \nu^{-4}$ or $\tau_{\rm sc}\propto \nu^{-4.4}$ \cite{Williamson1972,Bhat2004,Luan2014,Cordes2016,Farah2018,Xu2016,Qiu2020}. These frequency-dependent scattering are universal in observations, a propagation effect that cannot be ignored. 
Scattering causes a random interference pattern on the radio burst known as scintillation \cite{Rickett1977}, which manifests as frequency-dependent intensity modulation characterized by the decorrelation bandwidth $\Delta\nu$, specified at a certain frequency. Interference will happen when $2\pi\Delta\nu\tau_{\rm sc}\sim 1$, the decorrelation bandwidth widens with increasing frequency $\Delta\nu\sim 1/{2\pi}{\tau_{\rm sc}}$.

\subsection{Dispersion measure}
Different frequencies of electromagnetic waves travel at slightly different speeds due to their interaction with free electrons in the medium, which will cause different arrival times. For a typical plasma medium, the dispersion relation satisfies the following equation,
\begin{equation}
    \omega^2 = \omega_p^2 + k^2c^2,
\end{equation}
$\omega_p=\sqrt{4\pi n_e e^2/m_e}$
where $\omega$ is the frequency of electromagnetic waves, $\omega_p=\sqrt{4\pi n_e e^2/m_e}$, $n_e$ is the electron number density, $m_e$ is the mass of electron, $k$ is the wave vector and $c$ is the speed of light. For electromagnetic waves propagating in a plasma medium, the group velocity satisfies, 
\begin{equation}
    v_g = \frac{d\omega}{dk}=c\sqrt{1-(\omega_p/\omega)^2}.
\end{equation}
The time delay between two frequencies $\nu_1$ and $\nu_2$ caused by dispersion effect is, 
\begin{equation}
    \delta t = \frac{e^2}{2\pi m_e c}\left(\frac{1}{v_1^2}-\frac{1}{v_2^2}\right)\int n_e dl.
\end{equation}
Considering the evolution of redshift, the dispersion measure (DM) in the observer's frame is 
\begin{equation}
    {\rm DM}=\int\frac{n_e}{1+z}dl,
\end{equation}\label{eq:DM}
where $z$ is the redshift. DM is defined as the integral of free electron number density along the propagation path on the light of sight. 
The observed dispersion measure of most FRBs exceeds the corresponding contribution from the Milky Way, indicating that they likely originate from outside the Milky Way. 
The cosmological origin of FRBs makes them reliable probes for detecting the unknown electron number density in the universe. Additionally, the dispersion measure contributed from the intergalactic medium also reflects the distance of the source.

\subsection{Faraday rotation}
The orientation of the polarization plane of electromagnetic waves changes as they propagate through a magnetized plasma or magnetic field, called Faraday rotation.  
This results in a wavelength-dependent linear polarization,
\begin{equation}
    \Phi = {\rm RM}\lambda^2,
\end{equation}
where $\lambda$ is the wavelength. The definition of RM is
\begin{equation}
    {\rm RM}=0.81\int^d_0B_{//}n_e(l)dl,
\end{equation}
where $B_{//}$ is the magnetic field strength along the line-of-sight, $n_e$ is the number density of the medium and $d$ is the distance to the source. 
FRBs exhibit a wide range of measured RM values, from zero to several $10^5\,{\rm ram\,m^{-2}}$ \cite{Caleb2018,Michilli2018,Day2020}. FRB 20121102A, the first-discovered repeating FRB, shows a high RM value of $\sim 10^5\,{\rm rad\,m^{-2}}$ \cite{Michilli2018}. 
The relationship between RM and DM can be used to estimate the magnetic field strength along the line-of-sight \cite{WangWY2020,Lu2023}. 

Lu et al\cite{Lu2023} measured the strength of a magnetic field parallel to the line of sight using the simultaneous variation of rotation measure and dispersion measure of FRB 20201124A and found the strength ranges from a few $\mu$G to $10^3\mu$G. 
The absolute value of the average magnetic field $|\langle B_{//}\rangle|$ can be estimated as \cite{Xu2022,Katz2021}
\begin{equation}\label{eq_B}
    |\langle B_{//}\rangle| = 1.23\frac{|\Delta {\rm RM}|}{|\Delta {\rm DM}|}\,\mu{\rm G},
\end{equation}
The $|\Delta {\rm RM}|$ and $|\Delta {\rm DM}|$ represent the daily deviations from the average RM and DM values. Equation (\ref{eq_B}) is only valid when $\Delta B/B\ll\Delta {\rm RM}/{\rm RM}$ \cite{YangYP2023}.
The magnetic field of the host galaxies of FRBs is poorly constrained. The relation between the DM and the RM can be used to estimate the parallel magnetic field component of the host galaxy \cite{LinWL2016,Mannings2023}. 

The observed RM values of some repeating FRBs show variations \cite{Michilli2018,LuoR2020,Xu2022,NiuCH2022,Anna-Thomas2023}. The RM of active repeating FRB 20201124A exhibited irregular RM variations during its first 36-day active period, followed by a stable phase \cite{Xu2022}. It shows a local RM reversal \cite{WangFY2022}. Another active repeating FRB 20190520B showed an RM reversal on the order of $10^4$\,${\rm rad\,m^{-2}}$ \cite{Anna-Thomas2023}.

\subsection{Motivations behind this review}
This review focuses on the statistical analysis and cosmological applications of FRBs. As observational data accumulates, the classification and statistical analysis of FRBs have become increasingly important topics in the field of FRBs. The origin and physical mechanism of FRBs are still unknown, and statistical analysis provides necessary clues for exploring them. 
Whether all FRBs repeat is still a controversial issue and the comparison of these two populations helps to answer this question. The energy and waiting time distributions help us understand the origin and trigger mechanism of FRBs. Research on the properties of the FRB population is necessary for large sample data. In Section \ref{sec:statistic}, we present the studies of repetition, energy functions, waiting time distributions, and FRB populations, respectively.  

The cosmological origin makes FRB a reliable probe for cosmology. 
Newly constructed and commissioned instruments with enhanced sensitivity are expected to localize numerous FRBs to their host galaxies, thereby enhancing the cosmological utility of FRBs. In Section \ref{sec:cosmo}, we introduce the ``missing baryons" problem, measurements of cosmological parameters and the circumgalactic medium, gravitational lensing of FRB as a cosmological probe. 

We conclude the challenges and future prospects in the field of FRBs in Section \ref{sec:conclusion}.

\section{The statistical properties of FRBs}\label{sec:statistic}

\subsection{Non-repeating and repeating FRBs}
Up to now, more than 700 FRBs have been discovered, with a small fraction ($\sim$2.6\%) of them found to exhibit repeating pulses \cite{CHIME/FRB Collaboration2021,CHIME/FRB Collaboration2023}. 
The ones with repeating pulses are called repeating FRBs, while those that have not been found to repeat are called apparently non-repeating FRBs. We refer to them as non-repeating FRBs in the following discussion.
The most significant difference between repeating and non-repeating FRBs is their repetition rates. In reality, there is still debate about whether all FRBs are repeating \cite{Palaniswamy2018,Caleb2019,Ai2021}. 

In 2021, the CHIME/FRB Collaboration released the first CHIME/FRB FRB catalog, including bursts from repeaters and non-repeaters \cite{CHIME/FRB Collaboration2021}. This large sample provides the first opportunity to compare repeaters and non-repeaters. They found that repeaters and non-repeaters have the same sky, DM, scattering, flux, and fluence distributions utilizing the Anderson-Darling (AD) test and Kolmogorov-Smirnov (K-S) test. 
However, bursts from repeaters differ from non-repeaters in the intrinsic widths and bandwidths. 

\begin{figure*}
\centering
\includegraphics[width=\linewidth]{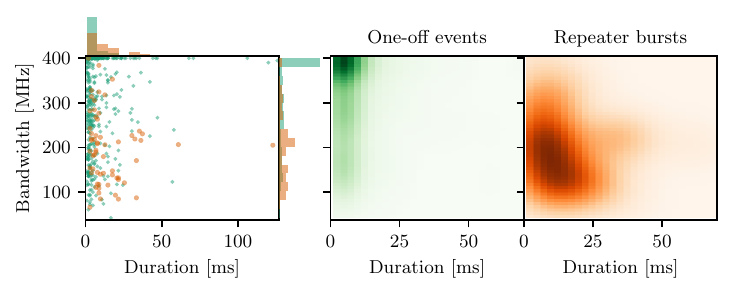}
\caption{{\bf Bandwidths and durations of the FRBs in the first CHIME/FRB FRB catalog with normalized histograms on the sides \cite{Pleunis2021b}.} FRBs are separated into non-repeater bursts (green diamonds) and repeater bursts (orange open circles). The panels on the right show smoothed and normalized distributions of all one-off events (green) and all repeater bursts (orange), respectively. Only bursts with detection S/N$>$12 are included. }
\label{fig:cat_duration_bw}
\end{figure*}

Using the 62 bursts from 18 repeaters and 474 non-repeaters from the first CHIME/FRB FRB catalog, Pleunis et al\cite{Pleunis2021b} confirmed that bursts from repeaters have larger widths and narrower bandwidths than non-repeaters. The statistical comparisons are shown in Figure \ref{fig:cat_duration_bw}.


As the number of repeating FRBs increases, the CHIME/FRB Collaboration conducted a statistical comparison including 25 newly-discovered repeaters between 2019 September 30 and 2021 May 1 \cite{CHIME/FRB Collaboration2023} and all other published CHIME/FRB-discovered repeaters \cite{CHIME/FRB Collaboration2019a,CHIME/FRB Collaboration2019b,KumarP2019,Fonseca2020,Bhardwaj2021,Lanman2022}. 
The same width and bandwidth distinctions between repeaters and non-repeaters have been drawn with the larger sample of repeaters. Furthermore, a significant difference in the DM distribution between repeaters and non-repeaters has been detected, which is not found in previous work. The DMs of repeaters are lower than those of non-repeaters. 
Zhong et al\cite{Zhong2022} performed a statistical analysis using the first CHIME/FRB FRB catalog and identified the repeaters and non-repeaters were two populations. 
Machine learning methods are used to classify repeaters and non-repeaters, and two apparent populations are found \cite{ChenBH2022,Luo2023,Zhu-Ge2023}.

Kirsten et al\cite{Kirsten2024} reported the detection of 46 high-energy bursts of FRB 20201124A using four 25-32 m class radio telescopes and they found a high-energy burst distribution that resembles that of the non-repeating FRB population. Based on this, they suggest that non-repeating FRBs may be the rarest bursts from repeating sources. Ai et al\cite{Ai2021} also provided similar discussions.

There is no definitive observational evidence to indicate that current non-repeating bursts will not repeat. One important argument that most FRBs should not be non-repeaters was the event rate, i.e. the rate density of FRBs already exceeded known abundant catastrophic events such as supernovae \cite{Ravi2019NA,LuoR2020}.
Whether all FRBs are repeating remains a mystery. 
Detailed population synthesis studies and multi-wavelength follow-ups are necessary to study the similarities and differences between repeaters and current non-repeaters.

\subsection{Energy functions}
Energy is crucial for understanding the physical origin and radiation mechanisms of FRBs. The energy of a FRB pulse can be calculated as  
\begin{equation}
    E = \frac{4\pi d_L^2 F\Delta\nu}{1+z},
\end{equation}
where $d_L$ is the luminosity distance of FRB, $F$ is the fluence, $z$ is the redshift, and $\Delta\nu$ is the bandwidth of the pulse. 
Previous works calculate the energy with central frequency $\nu_c$ \cite{ZhangB2018,LiD2021}. 
Zhang (2023) proposed either $\Delta\nu$ or $\nu_0$ could be used depending on whether the spectrum is narrow or broad \cite{ZhangB2023}. The bandwidth $\Delta\nu$ used here because many repeating bursts have narrow spectra.

The energy distribution of FRBs has always been a hot topic in the field of FRBs, with previous studies limited by data samples that only allowed analysis of a few repeating bursts, such as the first-discovered repeater FRB 20121102A \cite{Lu2016,Caleb2016,Law2017,WangFY2017,Gajjar2018,LuoR2018,WangFY2019,ZhangGQ2019,Lu2019,Cheng2020,Lu2020,LuoR2020}. With telescopes like CHIME, ASKAP, and FAST detecting increasing amounts of data, statistical analysis of energy distribution in large samples is becoming feasible \cite{Lu2020b,ZhangR2021,ZhangGQ2021, Hashimoto2022,Lyu2022}. 

A power-law distribution is widely used to fit the energy distribution of FRBs, 
\begin{equation}
    dN/dE \propto E^{-\alpha_E},
\end{equation}
where $\alpha_E$ is the slope of the power-law distribution. 

For individual repeating FRB, FRB 20121102A is the first FRB discovered by the Arecibo telescope to exhibit repeating bursts \cite{Spitler2016}. Over 12 years of long-term monitoring conducted by various radio telescopes, it has accumulated thousands of bursts.  
Wang \& Yu\cite{WangFY2017} found the energy distribution is $dN/dE\propto E^{-1.8}$ using 17 bursts of FRB 20121102A. Law et al\cite{Law2017} gave a similar energy distribution $dN/dE\propto E^{-1.7}$ with the VLA, Arecibo, and the Green Bank Telescope (GBT) bursts. A steeper slope $\alpha_E=2.8$ was found by Gourdji et al\cite{Gourdji2019} using 41 pulses of FRB 20121102A observed by the Arecibo telescope at 1.4GHz. Wang \& Zhang\cite{WangFY2019} performed a universal energy distribution with slope $\alpha_E=1.6-1.8$ of FRB 20121102A by different telescopes at different frequencies. 
Oostrum et al\cite{Oostrum2020} suggested a slope of $2.7\pm 0.6$ above the completeness threshold using the Apertif data. An index $2.1\pm 0.1$ was found by Cruces et al\cite{Cruces2021} using the Effelsberg data. The energy functions of FRB 20121102A remain controversial and require more data spanning a large energy range for a better understanding. 

FAST conducted continuous monitoring of FRB 20121102A from 29 August to 29 October 2019 \cite{LiD2021}. And 1652 bursts have been detected in a total of 59.5\,h, which is the largest sample of repeating FRBs. Providing research data for studying the physical nature and central engine of repeating FRBs, eliminating selection biases introduced by different telescopes. 
Utilizing these data, Li et al\cite{LiD2021} fitted the energy distribution of FRB 20121102A with a log-normal distribution plus a generalized Cauchy function:
\begin{equation}
N(E) = \frac{N_0}{\sqrt{2\pi}\sigma_E E} \exp\left[\frac{-(\log{E}-\log{E_0})^2}{2\sigma_E^2}\right] +
\frac{\epsilon_{E}}{1+(E / E_0)^{\alpha_E}},
\end{equation}
where $\epsilon_{E} = 0$ for $E<10^{38}$ erg and $\epsilon_{E} = 1$ for $E>10^{38}$ erg. The characteristic energy $E_0$ is $4.8\times 10^{37}$~erg. The best-fit distribution index is $\alpha_E = 1.85\pm0.30$. A power-law distribution could describe the high energy data, which is consistent with the previous work \cite{Gourdji2019}. However, the full energy data could not be well fitted by a single power-law and the real energy distribution is bimodal. Zhang et al\cite{ZhangGQ2021} used the same sample and proposed the energy count distribution $dN/dE$ at the high-energy range ($>10^{38}$\,erg) could be fitted by a single power-law function with an index of $1.86\pm 0.02$, while the distribution at the low-energy range ($<10^{38}$\,erg) deviates from the power-law function. 

\begin{figure}[!htp]
\centering
\includegraphics[width=\linewidth]{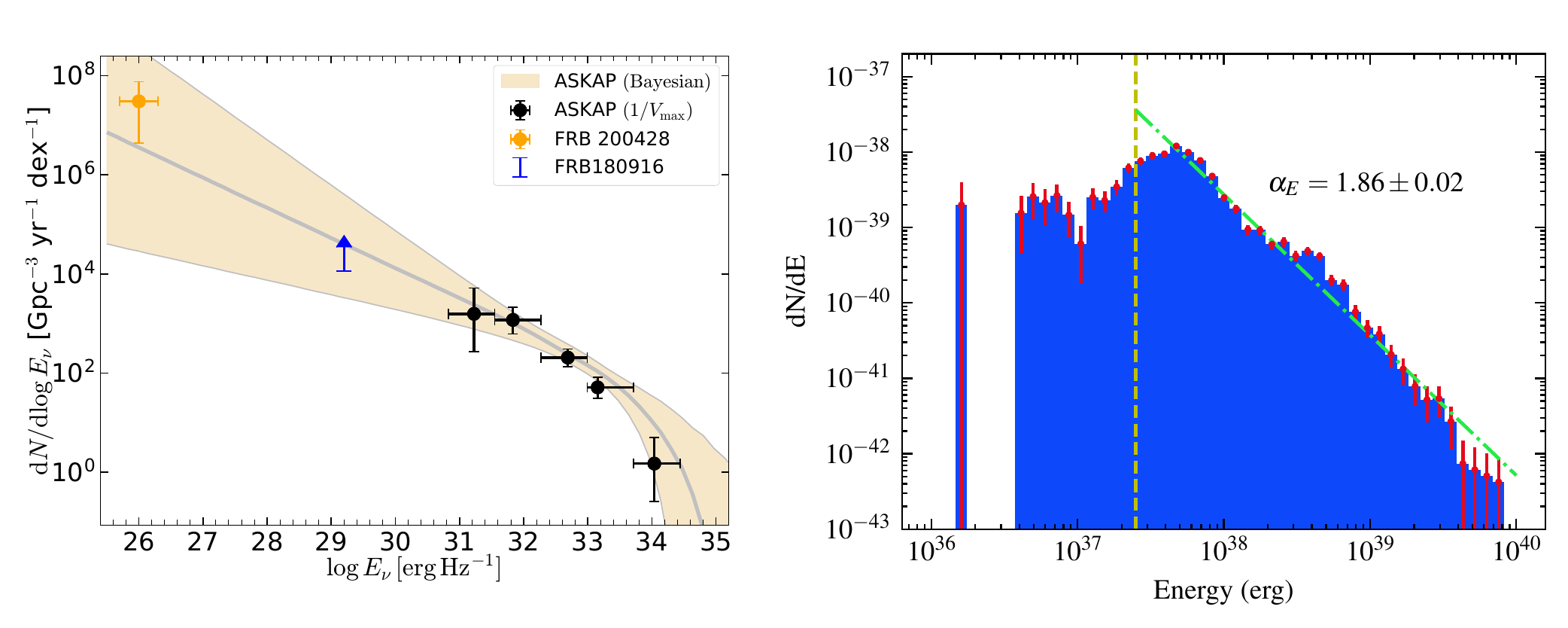}
\caption{{\bf Left panel: Energy distribution of different FRBs \cite{Lu2020}.} The volumetric rate at the faint end as inferred from FRB 20200428 (orange point with 68 percent C.L. Poisson errors), as compared to the rate at the bright end inferred from the ASKAP sample (the shaded region). The silver lines mark the 16 ($-1 \sigma$), 50 (median), and 84 percent ($+1 \sigma$) percentiles based on the Bayesian posterior shown in fig 4 of Lu \& Piro\cite{Lu2019} and evaluated at redshift $z=0.3$. The blue arrow shows the 90 percent C.L. lower limit for the contribution to the total volumetric rate density by FRB 20180916 \cite{CHIME/FRB Collaboration2021}. {\bf Right panel: The differential energy distribution for the bursts of FRB 20121102A observed by FAST \cite{ZhangGQ2021}.} The blue histogram is the differential energy distribution and the red points and red vertical lines are the values and $1\sigma$ uncertainties. The vertical yellow dashed line is the 90\% completeness threshold of the FAST telescope Li et al \cite{LiD2021}. The simple power law is used to fit the energy distribution in the high-energy range ($E>10^{38}$ erg) and the best-fitting result is shown as the dot-dashed green line. The power-law index is $-1.86\pm0.02$. The bursts in the low energies deviate from the power-law form.}
\label{fig:energy}
\end{figure}

For non-repeaters, Lu \& Kumar\cite{Lu2016} found a slope ranging from 1.5 to 2.2 assuming all FRBs satisfy a universal energy distribution. Luo et al\cite{LuoR2018} provided a power-law index in the range of [1.8, 1.2]. Lu \& Piro\cite{Lu2019} found the energy distribution slope is $1.6\pm 0.3$ using the ASKAP FRBs. As shown in the left panel of Figure \ref{fig:energy}, the energy distribution among different FRB sources shows a power-law distribution with a slope of 1.8, which can be extended FRB all the way down to the energy of FRB 20200428 \cite{Lu2020}.

A broken power-law function was also used to model the cumulative energy distribution of some repeating FRBs.
\begin{equation} \label{eq:broke}
N(\geq E) \propto \begin{cases}E^{-\alpha_{E1}} &  \quad \mathrm{for}~~ E<E_b \\ E^{-\alpha_{E2}} &  \quad \mathrm{for}~~ E\geq E_b\end{cases},
\end{equation}
in which $\alpha_{E1}$ and $\alpha_{E2}$ are the power-law indices for weak and strong bursts, and $E_b$ is the break energy.
Aggarwal et al\cite{Aggarwal2021a} suggested that the cumulative burst energy distribution of FRB 20121102A detected by Arecibo at 1.4\,GHz exhibits a broken power-law shape, with the lower and higher-energy slopes of $0.4\pm 0.1$ and $1.8 \pm 0.2$, with the break at $(2.3 \pm 0.2)\times 10^{37}$\,erg. 
Hewitt et al\cite{Hewitt2022} reported the energy distribution of a burst storm from FRB 20121102A in 2016 could be well-fitted by a break power-law function with $E_b=1.15\pm0.04\times10^{38}$\,erg, $\alpha_{E1}=1.38\pm0.01$, and $\alpha_{E2}=1.04\pm0.02$. A similar analysis with the FRB 20121102A rain in 2018 has been reported by Jahns et al\cite{Jahns2023}. 

Other repeating FRBs \cite{Xu2022,ZhouDJ2022,ZhangYK2023}, such as FRB 20201124A, have accumulated thousands of bursts.  
FRB 20201124A is an active repeating FRB first discovered by CHIME \cite{Lanman2022}. Lanman et al\cite{Lanman2022} performed a power-law function with the best-fit index $\alpha_E=4.6\pm1.3\pm0.6$ to estimate the CHIME/FRB-detected bursts of FRB 20201124A in 2021. Marthi et al\cite{Marthi2022} gave an index $\alpha_E=1.2\pm0.2$ using the bursts detected by the upgraded Giant Metrewave Radio Telescope (uGMRT) at 550-750 MHz. 
Using the 1863 bursts detected by FAST in 2021, Xu et al\cite{Xu2022} gave a broken power-law fitting with $\alpha_{E1}=0.36\pm0.02\pm0.02$, $\alpha_{E2}=1.5\pm0.1\pm0.1$ and $E_b=1.1\pm0.1\pm0.1\times 10^{38}$\,erg for the energy distribution of FRB 20201124A. While FAST caught 542 bursts on UTC 2021 September 25-28, Zhang el al.\cite{ZhangYK2022} gave the lower and higher-energy indices $\alpha_{E1}=1.22\pm 0.01$ and $\alpha_{E2}=4.27\pm0.23$ for the energy distribution fitting, respectively. 

The number of non-repeating FRBs is much larger than repeating FRBs and constraining the energy functions of them is particularly illuminating for better understanding FRB origins. To reflect the accurate properties of the intrinsic energy distribution of FRBs as much as possible, statistical analysis should be conducted using data observed from the same telescope to correct selection effects. 
Luo et al (2018)\cite{LuoR2018} and Luo et al (2020)\cite{LuoR2020} used the cutoff power-law function to account for the energy/luminosity function of the whole non-repeating FRB population.
James et al\cite{James2022a} modeled the FRBs detected by ASKAP and Parkes with a power-law with a maximum energy cutoff. The estimated maximum FRB energy is ${\rm log}_{10}E_{\rm max}({\rm erg}) = 41.70^{+0.53}_{-0.06}$ and the slope is $-1.09^{+0.14}_{-0.10}$.  

The cutoff power-law function is also used to fit the energy functions of FRBs,
\begin{equation}
    \phi({\rm log}E){\rm d log}E = \phi^*\left(\frac{E}{E^*}\right)^{\alpha+1}{\rm exp}\left(-\frac{E}{E^*}\right){\rm d log}E,
\end{equation}
where $\phi^*$ is the normalization factor, $\alpha+1$ is the faint-end slope, and $E^*$ is the break energy of the Schechter functions. Hashimoto et al\cite{Hashimoto2022} found the energy functions of non-repeating FRBs Schechter function-like shapes at $z<1$ using a homogeneous sample of 164 non-repeating FRB sources from the first CHIME/FRB FRB catalog. 
Shin et al\cite{Shin2023} found a slope of $\alpha=-1.3^{+0.7}_{-0.4}$ using 536 FRBs from the first CHIME/FRB FRB catalog, which is in broad agreement with Hashimoto et al\cite{Hashimoto2022}. 
Recent works \cite{Chen2024,ZhangKJ2024} used the Lynden-Bell's method \cite{Lynden-Bell1971} to study the luminosity/energy functions of FRBs from the first CHIME/FRB catalog. 

Although the power-law slope varies among different sources and across different observing epochs, Kirsten et al\cite{Kirsten2024} found that it is still apparent that the slopes of non-repeating FRBs exhibit a flatter trend compared to repeating FRBs. They compared the energy distributions of low-energy and high-energy bursts from FRB 20201124A and found a possible link between repeaters and non-repeaters. 
This encourages more radio telescopes to engage in long-term monitoring and multi-wavelength observations of FRBs in the future.

Similar power-law energy distributions have been discovered in a large number of astrophysical phenomena, such as hard X-ray and gamma-ray flares from the Sun \cite{Aschwanden2015}, soft gamma-ray repeaters \cite{Cheng1996,Cheng2020}, and gamma-ray bursts \cite{WangFY2013}. 
One concept to explain and model these observed power-law distributions is the concept of self-organized criticality (SOC) in slowly driven nonlinear dissipative systems proposed by Bak et al\cite{Bak1987} and Katz\cite{Katz1986}. 
In a SOC system, due to some driving force, subsystems will self-organize to a critical state at which a small perturbation triggers an avalanche-like chain reaction within the system \cite{Bak1987}. 
SOC theory also predicts that the waiting time distributions will show a power-law function or exponential function, depending on the form of the event rate. Below, we will discuss the waiting time distribution of FRBs.

\subsection{Waiting time distributions}

Waiting time $\Delta T$ is defined as the time interval between two adjacent bursts, 
\begin{equation}
    \Delta T = T_{\rm i+1} - T_{\rm i},
\end{equation}
where $T_{\rm i+1}$ and $T_{\rm i}$ are the arrival time of two adjacent bursts. 
The waiting time for repeating FRBs spans from milliseconds to several thousand seconds. 
Unlike pulsars and Rotating Radio Transients (RRATs), the activity of FRBs is intermittent and irregular \cite{Spitler2016}. The repetition of FRBs manifests in the distribution of waiting times, providing crucial clues for understanding the central engines and emission mechanisms of repeating FRBs \cite{ZhangB2020,LiD2021}.

\subsubsection{The cumulative distribution}

\begin{figure}[!htp]
\centering
\includegraphics[width=\linewidth]{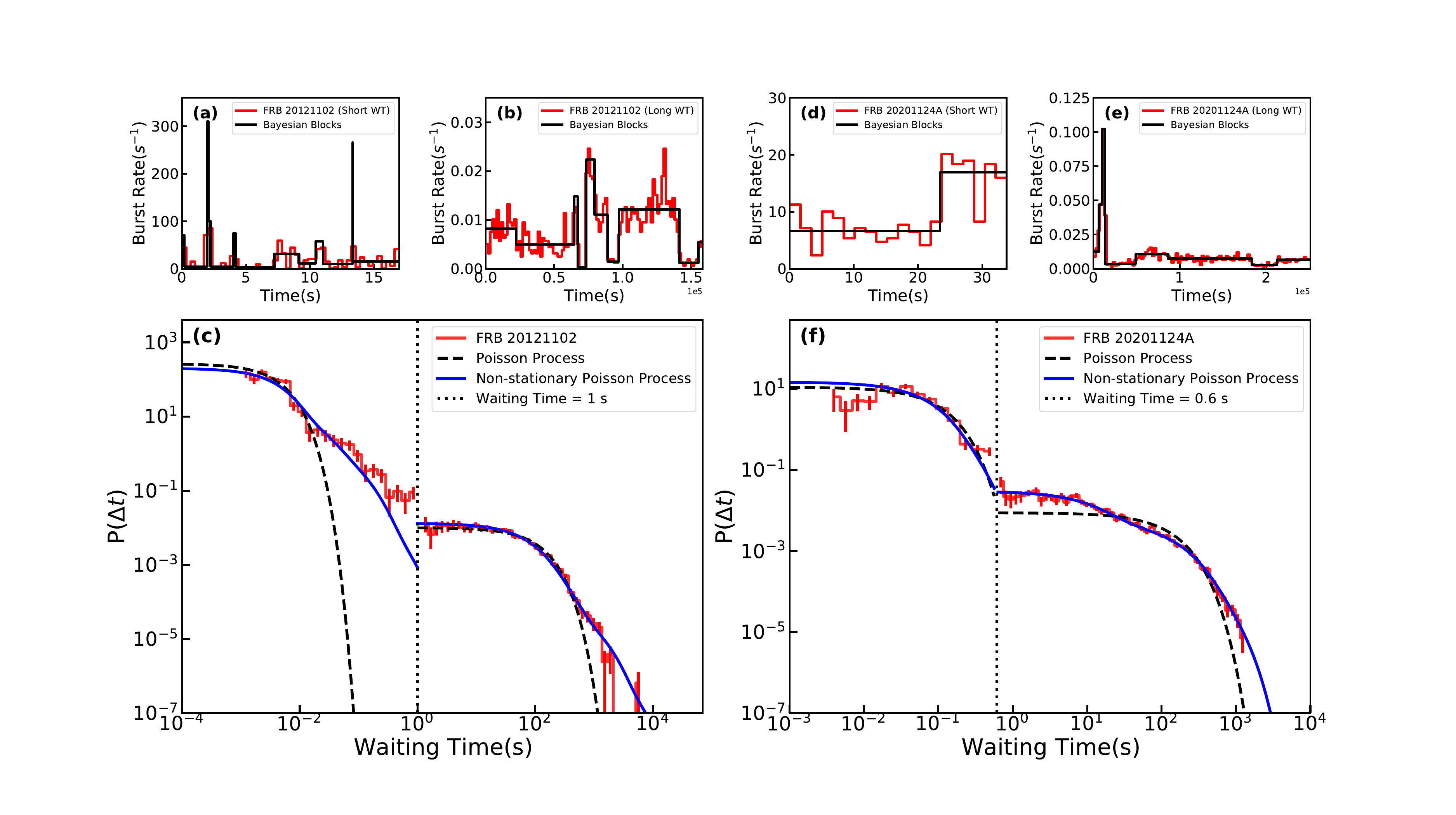}
\caption{{\bf The cumulative waiting time distribution of repeating FRB 20121102A and FRB 20201124A \cite{WangFY2023}.} 
Panel (a). Bayesian blocks decomposition of the burst rate for short waiting times ($\Delta t<1$ s) of FRB 20121102A. 
Panel (b). Bayesian blocks decomposition of the burst rate for long waiting times ($\Delta t>1$ s) of FRB 20121102A. 
Panel (c). The red stepwise line is the observed distribution of waiting time of FRB 20121102A. The solid lines show the fitting result for a non-stationary Poisson process with a distribution of rates estimated from the data (panels (a) and (b)). The dashed lines give the fitting results of a Poisson process. From the reduced $\chi^2$, we can see that the distribution of long waiting times can be explained by a non-stationary Poisson process. Panel (d). Same as panel (a), but for FRB 20201124A. Panel (e). Same as Panel (b), but for FRB 20201124A. Panel (f). Same as panel (c), but for FRB 20201124A. The distribution of long waiting times for FRB 20201124A can be well explained by a non-stationary Poisson process.}
\label{fig:waiting2}
\end{figure}

The study of waiting time distributions has gradually improved with the increasing data on repeating FRBs. In the beginning, the cumulative distribution of waiting time was first studied due to insufficient data.  
As the first discovered repeating FRB, FRB 20121102A has the longest history of research on its waiting time cumulative distribution.
Wang \& Yu (2017) found that the waiting time distribution of FRB 20121102A is non-Poisson \cite{WangFY2017}. 
For a time series with a period $P$, the waiting time distribution will accumulate at multiples of the period or at the period itself. For a time series of completely random events, the distribution of waiting times follows an exponential distribution. 
For such a Poisson process, its waiting time distribution $P(\Delta t)$ can be written as
\begin{equation}
    P(\Delta t) = \lambda e^{-\lambda\Delta t},
\end{equation}
where the constant $\lambda$ is the burst rate of events \cite{Wheatland1998}. 
Non-stationary distribution \cite{WangFY2017,Cheng2020,WangFY2021,Jahns2023,WangFY2023,WangP2023} is used to fit the waiting time distribution of FRBs. 
The non-stationary process can be considered a superposition of multiple stationary processes with varying occurrence rates \cite{Cheng2020}.
\begin{equation}
P(\Delta t)=\left\{\begin{array}{ll}{\lambda_{1} e^{-\lambda_{1} \Delta t}} & {\text { for } t_{1} \leq t \leq t_{2}} \\ {\lambda_{2} e^{-\lambda_{2} \Delta t}} & {\text { for } t_{2} \leq t \leq t_{3}} \\ {\ldots \ldots \ldots} & {\text { for } t_{n} \leq t \leq t_{n+1}} \\ {\lambda_{n} e^{-\lambda_{n} \Delta t}} & {\text { for } t_{n} \leq t \leq t_{n+1}}\end{array}\right.
\end{equation}
For the time-dependent burst rate $\lambda(t)$, it can be written as a combination of piecewise constant Poisson processes \cite{Aschwanden2010} 
\begin{equation}
    P(\Delta t)=\sum_i\phi_i\lambda_i {\rm exp}(-\lambda_i\Delta t).
\end{equation}
This formula can be expressed in the form of a power law under certain conditions.
Wang \& Yu\cite{WangFY2017} found the waiting time distribution shows the power-law distribution, and is consistent with a non-stationary Poisson process. 
Wang et al\cite{WangFY2023} used the Bayesian blocks method to fit the waiting time distribution of FRB 20121102A and FRB 20201124A, and they found they exhibit an approximate power-law tail, which is consistent with a Poisson model with a time-varying rate. 
The cumulative distributions of FRBs and the non-stationary function fittings are shown in Figure \ref{fig:waiting2}. Their findings support a correlated trigger for FRBs. 

Some works suggest that the waiting time distribution of FRBs follows the Weibull distribution \cite{Oppermann2018,Oostrum2020,ZhangGQ2021,Cruces2021}, which is 
\begin{equation}
\label{eq:weibull}
\mathcal{W}(\delta_t \mid k, r)=k \delta_t^{-1}[\delta_t r \Gamma(1+1 / k)]^{k} \mathrm{e}^{-[\delta_t r \Gamma(1+1 / k)]^{k}},
\end{equation}
where $\delta_t$ is the waiting time, $k$ is the shape parameter, $r$ is the mean burst rate, and $\Gamma$ is the gamma function.
The case $k = 1$ corresponds to the Poisson distribution. Previous works suggested that the repeating behavior of FRB 20121102A tends to have $k < 1$ \cite{Oppermann2018,Oostrum2020,Cruces2021,ZhangGQ2021}, which means that time clustering is favored. 

There is still controversy over whether the bursts of FRBs are correlated. Earthquakes are used to analogize with FRBs \cite{WangWY2018,Totani2023,Xie2024}. Totani \& Tsuzuki\cite{Totani2023} reported a clear power-law correlation function analysis of repeating FRBs in the time-energy two-dimensional space and they demonstrated that FRBs are common to earthquakes. 
However,  there are some studies with opposing views that suggest FRBs are different from seismic events.
Zhang et al\cite{ZhangYK2024} suggested that the emission of FRBs does not exhibit the time and energy clustering using ``Pincus Index" and "Maximum Lyapunov Exponent", which is different from lots of previous works \cite{Oppermann2018,Oostrum2020,Cruces2021,ZhangGQ2021,WangFY2017,Cheng2020,WangFY2021,Jahns2023,WangFY2023}.  
There are other models proposed to fit the waiting time of repeaters, such as the Bent power-law distribution \cite{LinHN2020,Sang2023}, the broken power-law distribution \cite{Panda2024}, and the Tsallis q-Gaussian distribution \cite{LinHN2020,WangFY2021,GaoCY2024,Du2024}. 
We have summarized some studies on the waiting time distribution in Table \ref{table:wt}. The statistical properties of repeating FRBs are similar to earthquakes, suggesting that the trigger mechanism of the correlated repeating bursts may come from the crustal failure mechanism of neutron stars, namely starquakes.

\begin{table}[t]
\begin{center}
\caption{Summary of Researches on Waiting Time Distribution}
\begin{tabular}{lccc} \hline
   Methods   & Physical Models & References  \\\hline
   Log-normal distribution fitting  & - & \cite{LiD2021,Xu2022,Hewitt2022,ZhangYK2023}  \\
   Weibull distribution fitting & Clustering/non-Poissonian nature & \cite{Oppermann2018,ZhangGQ2021,Cruces2021,Oostrum2020}
   \\
   Non-stationary Poisson distribution fitting  & Self-organized criticality & \cite{WangFY2017,Cheng2020,WangFY2021,Jahns2023,WangFY2023,WangP2023} \\
   Correlation function analyse & Earthquake-like origin & \cite{Tsuzuki2024,Totani2023}    \\
   Exponential distribution fitting & Brownian motion   & \cite{ZhangYK2024}   \\
   \hline
\end{tabular}
\label{table:wt}
\end{center}
\end{table}

\subsubsection{The frequency distribution}
\begin{figure}[!htp]
\centering
\includegraphics[width=\linewidth]{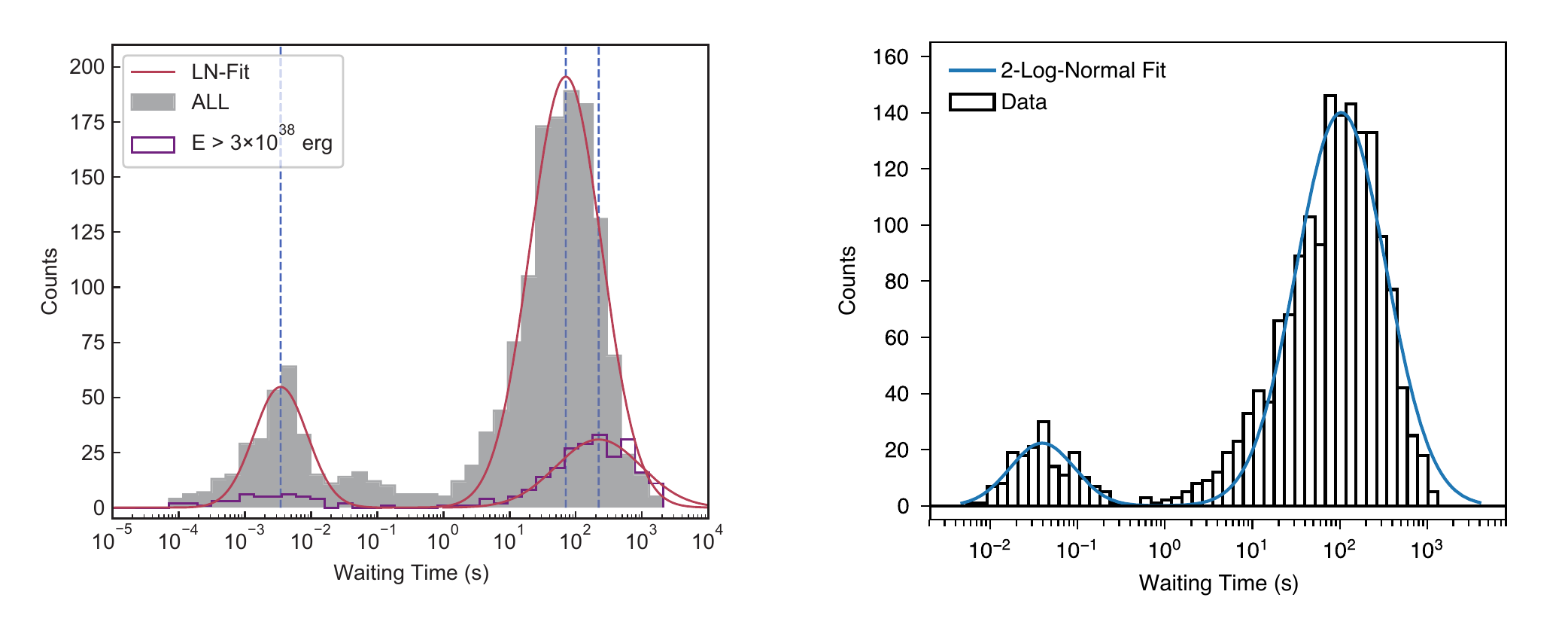}
\caption{{\bf Left panel: The waiting time distribution of FRB 20121102A \cite{LiD2021}.} 
The grey bar and solid red curve denote the distribution of waiting time and its log-normal (LN) fit. The high energy component ($E>3\times10^{38}$~erg) is shown as the solid purple line. The three fitted peak waiting times (blue dashed vertical lines) from left to right are $3.4\pm1.0$~ms, $70 \pm 12$~s, and $220\pm100$~s, respectively. The peaks around 70\,s and 220\,s in the waiting time distribution are close to the  average values for the respective samples (full and high energy). 
{\bf Right panel: The waiting time distribution of FRB 20201124A \cite{Xu2022}.} The best fit using two log-normal functions (the blue curve), where the two log-normal distributions peak at 39\,ms and 106.7\,s.}
\label{fig:waiting1}
\end{figure}

As a large amount of repeating FRB data has been observed, with FAST playing a crucial role, more data are now available for studying the waiting time distributions.
A bimodal distribution with millisecond and second peaks has been widely found in repeating FRBs \cite{Gourdji2019, LiD2021, Aggarwal2021,Cruces2021,LiB2019,WangYB2023,HuCR2023}, which shows non-Poisson clustering of bursts. 
FAST has made significant contributions in discovering a large number of repeated bursts. The FAST data firmly established the bimodal distribution thanks to the large amount of bursts detected.
Gourdji et al\cite{Gourdji2019} the waiting time distribution of 41 bursts from FRB 20121102A observed by the Arecibo telescope at 1.4\,GHz follows a log-normal distribution centered at $207\pm 1$\,s. Adding 93 bursts, Aggarwal et al\cite{Aggarwal2021} found that the peak of log-normal waiting time distribution of FRB 121102 decreases from $207\pm 1$\,s to $74.8\pm 0.1$\,s compared to that of Gourdji et al\cite{Gourdji2019}. The right peak of waiting time distribution is affected by the event rate in an observation of constant length. Independent groups have performed similar analyses of the waiting time distribution of FRB 20121102A \cite{ZhangYG2018,Katz2018,LiB2019}. 
FAST detected a large sample of FRB 20121102A \cite{LiD2021} and the waiting time can be well fitted by a log-normal function centered at $70\pm 12$\,s. The waiting time distribution is shown in the left panel of Figure \ref{fig:waiting1}. A similar peak ($\sim 95$\,s) has been found by Hewitt et al\cite{Hewitt2022}, who presented a burst storm from FRB 20121102A detected by Arecibo in 2016. 

Similar bimodal distributions of waiting time have also been observed in other repeaters \cite{Nimmo2023,CaiC2022,Xu2022,ZhangYK2022,Niu2022,ZhangYK2023}. The CHIME-discovered FRB 20201124A is an active repeater, which has a significant, irregular, short-time variation of RM during FAST observation windows in 2021 \cite{Xu2022}. Two log-normal functions with two peaks at 39\,ms and 106.7\,s can fit the waiting time distribution of FRB 20201124A \cite{Xu2022}, as shown in the right panel in Figure \ref{fig:waiting1}. While Zhang el al.\cite{ZhangYK2022} gave well-fitted two log-normal functions with peaks 51.22\,ms and 10.05\,s using 882 bursts detected by FAST in 2022, their lower right peak shows it depends on the activity level of the source. 
Like FRB 20121102A and FRB 20201124A, FRB 20200912A also exhibits a bimodal distribution with peaks around 51\,ms and 18\,s \cite{ZhangYK2023}. 
The M81 globular-associated repeating FRB 20200120E also shows a bimodal waiting time distribution with two peaks of $0.94^{+0.07}_{-0.06}$\,s and $23.61^{+3.06}_{-2.71}$\,s \cite{Nimmo2023}. 
It is widely believed that the right peak at the seconds is related to the activity of the source. Therefore, even though the right peak from the same source may vary at different observation periods, such as FRB 20201124. The left peak occurs in milliseconds and may be influenced by different pulse division criteria. Some points suggest that the location of this peak depends on the size of the emission region or propagation process \cite{Nimmo2023,Xiao2024}.

\subsection{Redshift distributions}
Due to the limited number of localized FRBs, our understanding of the redshift distribution of FRBs is currently very limited. The redshift of unlocalized FRBs can be inferred from the DM value. 
Locatelli et al\cite{Locatelli2019} applied \textlangle$V/V_{\rm max}$\textrangle test for the ASKAP and Parkes data. They suggested that the population of FRB progenitors evolves faster than the star formation rate, while the Parkes value is consistent with it. 
Hashimoto et al\cite{Hashimoto2020b} proposed that non-repeating FRBs follow the stellar-mass evolution of long-living objects and repeating FRBs do not evolve with redshift. 
Zhang et al\cite{ZhangRC2021} tested two redshift distribution models, one tracking the star formation history of the Universe and the other tracking compact binary mergers. They found that FRBs from the Parkes or ASKAP are compatible with both models. 
James et al\cite{James2022b} found the fast radio burst population evolves, consistent with the star formation rate using the ASKAP and Parkes data. 

The first CHIME/FRB catalog released by the CHIME collaboration significantly increased the number of FRBs, thereby strengthening the constraints on the redshift distribution of FRBs. 
Zhang et al\cite{ZhangRC2022} proposed that the CHIME FRB data are consistent with a redshift model that exhibits a delay relative to star formation. 
Hashimoto et al\cite{Hashimoto2022} suggested old populations including neutron stars and black holes, as more likely progenitors of non-repeating FRBs from the first CHIME/FRB catalog.
Qiang et al\cite{Qiang2022} concluded that the FRB distribution model requires a certain degree of "delay" with respect to the cosmic star formation history. 
However, Shin et al.\cite{Shin2023} found the CHIME data still follows the star formation history, which may be affected by the sample selection .
However, Shin et al.\cite{Shin2023} found the CHIME data still follows the star formation history, which may be affected by the selection criteria for high-SNR samples.
Recent works \cite{Chen2024,ZhangKJ2024} used the Lynden-Bell's method \cite{Lynden-Bell1971} to study the redshift distributions of FRBs from the first CHIME/FRB catalog. Chen et al\cite{Chen2024} proposed the old populations are closely related to the origins of FRBs.

\section{The cosmological applications of FRBs}\label{sec:cosmo}
\subsection{Searching ``missing baryons"} 

In a flat $\Lambda$ cold dark matter (CDM) universe, cosmic microwave background (CMB) and Big Band nucleosynthesis (BBN) observations give $\sim$ 95\% of its composition as dark energy(69\%) and dark matter(26\%), with the remaining $\sim$5\% as baryonic matter \cite{Cyburt2016,Planck2020}. However, there are $\sim$ 30\% of the cosmic baryonic matter appears to be missing in observations, which is called the ``missing baryons" problem \cite{Fukugita2004,Cen2006,Bregman2007,Shull2012,McQuinn2016}. 
Finding the `invisible' baryon content in the intergalactic medium and the circum-galactic medium is essential for studying the formation and evolution of the large-scale structure of the universe and providing key insights into understanding galaxy evolution. 

The primary methods to measure the baryonic matter, such as the absorption line spectroscopy \cite{Shull2012,Nicastro2018}, quasar spectroscopy \cite{Tripp2002,Tumlinson2011,Prochaska2011}, Sunyaev-Zel'dovich analyses \cite{Hojjati2017,de Graaff2019} and X-ray emission studies \cite{Eckert2015} have their limitations.  
The dispersion measure of extra-galactic fast radio bursts with identified redshift can be an independent probe to measure baryon content in the intergalactic medium \cite{McQuinn2014,Macquart2020}.
The ensemble of the dispersion measure $\rm DM_{obs}$ and redshift $z$ measurements provides basic information on the electron density along the propagation path. The observed dispersion measure ${\rm DM_{obs}}(z)$ is composed of multiple components 
\begin{equation}\label{DMobs}
{\rm{DM_{obs}}}(z) = {\rm DM_{MW,ISM}}+{\rm DM_{MW,halo}}+{\rm DM_{IGM}}(z)+\frac{{\rm DM_{host}}}{(1+z)},
\end{equation}
where ${\rm DM_{MW,ISM}}$ and ${\rm DM_{MW,halo}}$ are the contributions from the interstellar medium (ISM) and halo of the Milky Way (MW), ${\rm DM_{IGM}}(z)$ is the contribution from the intergalactic medium (IGM) and ${\rm DM_{host}}$ is the combination from the host galaxy halo and the local environment of the FRB, which is weighted by a factor $1/(1+z)$. 
Among them, only ${\rm DM_{IGM}}(z)$ has a strong correlation with redshift, and extracting it from ${\rm{DM_{obs}}}$ is the key issue to studying the IGM. The galactic electron-density model provides an estimation of ${\rm DM_{MW,ISM}}$ \cite{Cordes2002,Yao2017}. 

For a flat universe, the average value of ${\rm DM_{IGM}}$ can be expressed as \cite{Deng2014}\cite{Macquart2020}: 
\begin{equation}\label{DM_IGM}
    \langle {\rm DM_{IGM}}(z)\rangle = \frac{3c\Omega_bH_0}{8\pi G m_p}\int_{0}^{z_{\rm FRB}}\frac{f_{\rm IGM}(z)f_e(z)(1+z)}{\sqrt{\Omega_m(1+z)^3+ \Omega_{\Lambda}}}dz, 
\end{equation}
where $m_p$ is the proton mass, $H_0$ is the Hubble constant, the cosmological parameters $\Omega_b$, $\Omega_{\Lambda}$ and $\Omega_m$ represent the density of baryons, dark matter and dark energy in the universe, respectively. The electron fraction $f_e(z) = Y_H X_{e, H}(z) + \frac{1}{2} Y_{He}X_{e, He}(z)$, with hydrogen (H) fraction $Y_H  = 3/4$ and helium (He) fraction $Y_{He} = 1/4$. 
Assuming hydrogen and helium are completely ionized at $z<3$, the ionization fractions of intergalactic hydrogen and helium $X_{e,H} = X_{e,He} = 1$. $f_{\rm IGM}(z)$ shows the evolution of the baryon fraction in the IGM. 
The value of $f_{\rm IGM}$ at different redshift has not been determined yet. Numerical simulations and observations give a estimation that $f_{\rm IGM}\approx 0.9$ at $z\geq 1.5$ \cite{Meiksin2009} and $f_{\rm IGM}\approx 0.82$ at $z\leq 0.4$ \cite{Shull2012}. 

\begin{figure*}
\centering
\includegraphics[width=\linewidth]{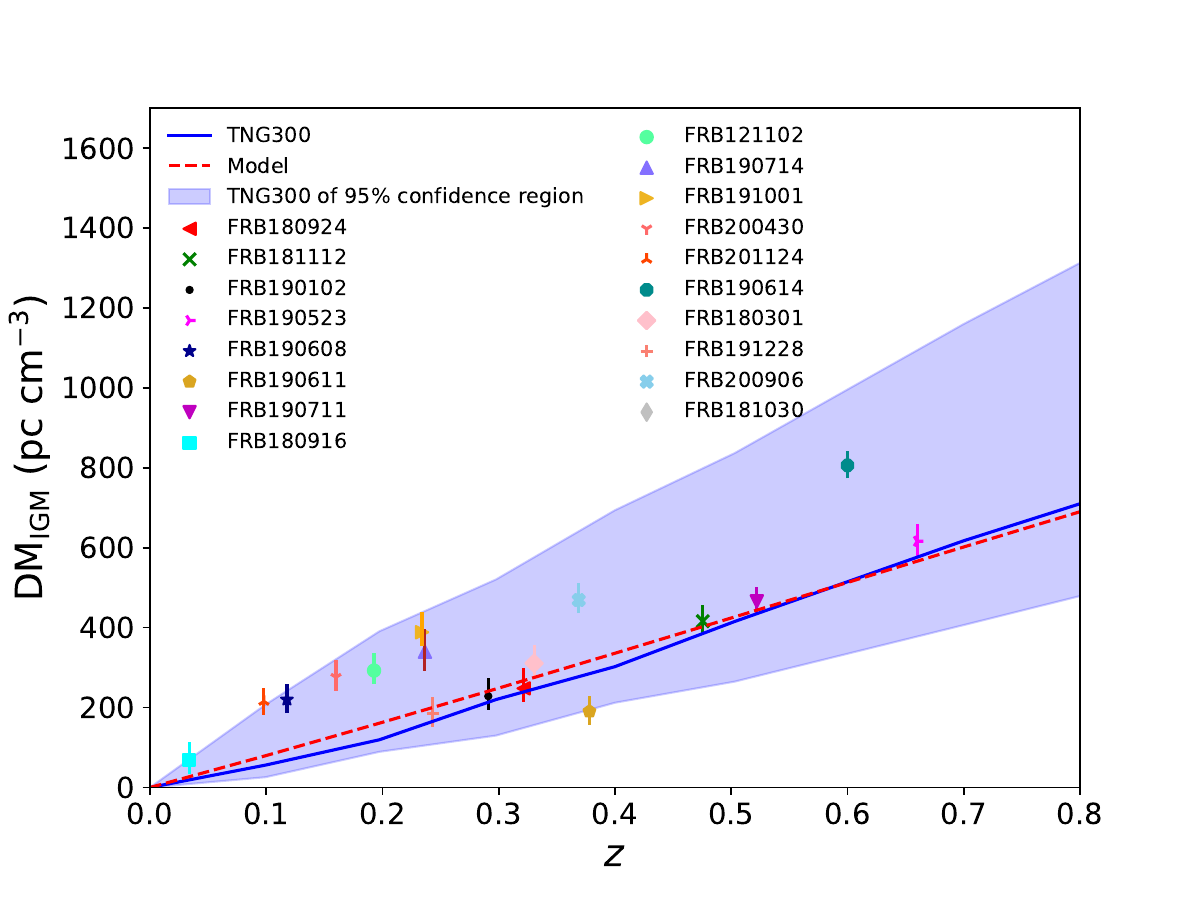}
\caption{{\bf The $ \rm \bf DM_{IGM}$-$\bm z$ relation for eighteen localized FRBs \cite{Wu2022}.} 
The scattered points are the $\rm DM_{IGM}$ values of the eighteen localized FRBs. The $\rm DM_{IGM}$ values are derived by correcting the observed dispersion measure $\rm DM_{obs}$ for the estimated contributions from our Galaxy and the host galaxy. The $\rm DM_{MW,ISM}$ is deduced from NE2001 model, and $\rm DM_{MW,halo}$ is adopted as 50 pc cm$^{-3}$. We use the median value of $\rm DM_{host}$ at different redshifts from IllustrisTNG 300 cosmological simulation.  The red dotted line shows model of equation (\ref{DM_IGM}) with $\Omega_{\rm m} = 0.315$, $\Omega_{\rm B}h^2 = 0.02235$ and $H_0 = {\rm 70\ km\ s^{-1}\ Mpc^{-1}}$. The blue line corresponds to the $\rm DM_{IGM}$ result from the IllustrisTNG 300 cosmological simulation and the purple shaded area is the 95\% confidence region. } 
\label{fig_dm_z}
\end{figure*}

Equation (\ref{DM_IGM}) gives a relation between the dispersion measure from IGM and redshift, which is called $\rm DM_{IGM}$-$z$ relation.
Figure \ref{fig_dm_z} shows the dispersion-redshift relation and the estimated ${\rm DM_{IGM}}$ of eighteen localized FRBs \cite{Wu2022}. This approximate linear dispersion measure-redshift relation has been found by several works\cite{Ioka2003,Inoue2004}, which is used to measure the baryonic mass density parameter $\Omega_{\rm b}$ \cite{Deng2014, McQuinn2014, Macquart2020, Yang2022} and the baryon fraction in the diffuse IGM ($f_{\rm IGM}$) \cite{LiZX2019,LiZX2020,Walters2019,Wei2019,Qiang2021,Dai2021,Lin2023,WangB2023}.

A key issue is to independently determine the DMs contributed by different parts during propagation. Here we discuss the contribution of each term according to Eq. (\ref{DMobs}).
The interstellar medium of the Milky Way contribution ${\rm DM_{MW,ISM}}$ can be derived from the galactic electron-density model \cite{Cordes2002,Yao2017}. 
The halo of the Milky Way contribution ${\rm DM_{MW,halo}}$ is not well constrained, and halo gas models can give an estimation of ${\rm DM_{MW,halo}}$ along the line of sight using cosmological simulations \cite{Dolag2015,Yamasaki2020,Prochaska2019}. The scatter of ${\rm DM_{MW,halo}}$ is lower than the uncertainty of the extragalactic DM contributions for the cosmological FRBs.

\begin{figure*}
\centering
\includegraphics[width=\linewidth]{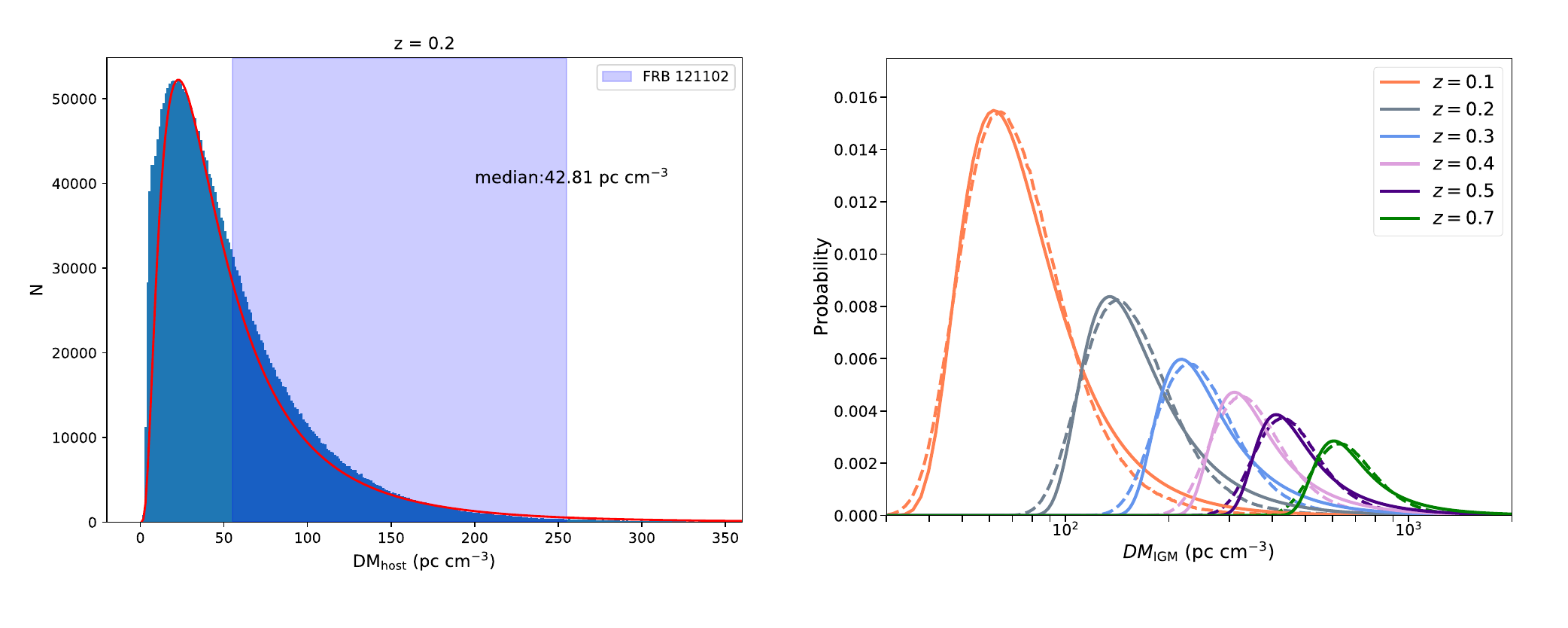}
\caption{{\bf Left panel: Distribution of $\rm DM_{host}$ at $z=0.2$ for repeating FRBs like FRB 121102 from the IllustrisTNG simulation \cite{Zhang2020}.} The red line is the best-fitting result using log-normal distribution. The blue-shaded region is the $\rm DM_{host}$ for FRB 121102 inferred from the observation.
{\bf Right panel: Distribution of $\rm DM_{IGM}$ from $z=0.1$ to 0.7 using the IllustrisTNG simulation \cite{Zhang2021}.} Dashed lines are $\rm DM_{IGM}$ distributions derived from IllustrisTNG simulations and solid lines are the fitting results using equation \ref{eq:DMIGM}. }
\label{fig_tng}
\end{figure*}

A more thorny problem is determining the contributions outside the Milky Way. The DM contribution from the host galaxy $\rm DM_{host}$ is hard to determine from observations. 
The galaxy type, morphology, and light-of-sight on the sky and offset of the FRB from the galaxy center will influence the $\rm DM_{host}$ value for an FRB. 
Cosmological simulations, such as IllustrisTNG simulation, will provide an effective method for modeling the $\rm DM_{host}$ distribution at different redshifts \cite{Macquart2020, Zhang2020}. 
The scatter of $\rm DM_{host}$ can be described using log-normal distribution \cite{Macquart2020},
\begin{equation}
p_{\rm host}({\rm DM_{host}}|\mu,\sigma_{\rm host}) = \frac{1}{(2 \pi)^{1/2} {\rm DM_{host}} \sigma_{\rm host}} \exp \left[ - \frac{(\log{\rm DM_{host}}-\mu)^2}{2 \sigma_{\rm host}^2} \right].
\label{eq:dmhost}
\end{equation}
Where $e^{\mu}$ and $e^{\mu +\sigma_{\rm host}^2/2} ( e^{\sigma_{\rm host}^2} - 1 )^{1/2}$ are the median and variance value of the distribution. Zhang et al\cite{Zhang2020} simulated the $\rm DM_{host}$ distribution from IllustrisTNG simulation as shown in the left panel of Figure \ref{fig_tng}. 

On the other hand, the existence of filaments, voids, and other structures in the universe leads to an inhomogeneous intergalactic medium. McQuinn\cite{McQuinn2014} has studied the inhomogeneous situation through simulations and found that the probability distribution of DM is related to the baryons. 
Jaroszynski\cite{Jaroszynski2019} has found a few percent scatters in DM to a source and their distribution of DM is close to Gaussian distribution. Macquart et al\cite{Macquart2020} gives a quasi-Gaussian function with a long tail to describe the $\rm DM_{IGM}$ distribution, 
\begin{eqnarray}
p_{\rm IGM}(\Delta) = A \Delta^{-\beta} \exp\left[ - \frac{(\Delta^{-\alpha} - C_0)^2}{2 \alpha^2 \sigma_{\rm DM}^2} \right], \qquad \Delta > 0,
\label{eq:DMIGM}
\end{eqnarray}
where $\Delta\equiv {\rm DM_{IGM}}/\langle{\rm DM_{IGM}}\rangle$, $\beta$ is related to the inner density profile of gas in halos, $\alpha$ is the index if the density profile scales as $\rho\propto r^{=\alpha}$ and $C_0$ is a constant. The cutoff low-${\rm DM_{IGM}}$ is contributed from the highly diffuse IGM while the long tail high-${\rm DM_{IGM}}$ is dominated by the halo-related component. 
Pol et al\cite{Pol2019} used the MareNostrum Instituto de Ciencias del Espacio Onion Universe simulation and estimated an intergalactic DM of ${\rm DM_{IGM}}(z=1)=800_{-170}^{+7000}\,\rm{pc\,cm^{-3}}$. 
Zhang et al\cite{Zhang2021} used the IllustrisTNG simulation to realistically estimate the $\rm DM_{IGM}$ distribution up to $z\sim 9$, and their results are shown in the right panel of Figure \ref{fig_tng}.

Giving the probability density distribution (PDF) of $\rm DM_{host}$ and $\rm DM_{IGM}$, for a localized FRB with redshift $z_{i}$ we have:
\begin{equation}
P_i({\rm DM_{\rm FRB, i}'| z_i})  =  \int\limits_0^{\rm DM_{\rm ex}} \;
  {p_{\rm host}({\rm DM_{host}}|\mu,\sigma_{\rm host}) } \; p_{\rm IGM}({\rm DM_{\rm ex, i} - {\rm DM_{host}}}, z_i) \; d{\rm DM_{host}}  .
\label{eq:prob}
\end{equation}
The extragalactic $\rm DM_{ex}$ is the combination of $\rm DM_{host}$ and $\rm DM_{IGM}$. 
For a sample of localized FRBs, the joint likelihood is 
\begin{equation}
 \mathcal{L} = \prod\limits_{i=1}^{N_{\rm FRBs}} P_i ({\rm DM_{ex}}|z_i), 
\label{eq:L}
\end{equation}

If one can distinguish each component of $\rm DM_{obs}$, one can directly measure the ``missing baryons'' in the universe through the $\rm DM_{IGM}-z$ relation.  
Due to the lack of localized FRBs, most precious works constrain the ``missing baryons'' through simulations \cite{McQuinn2014}. Macquart et al\cite{Macquart2020} derive a cosmic baryon density of $\Omega_{\rm b}=0.051^{+0.021}_{-0.025}$ with five arcsecond localized FRBs, which is consistent with values derived from CMB and BBN. 
As the number of localized FRBs increases, the statistical errors for $\Omega_{\rm b}$ become smaller \cite{Yang2022,Lin2023}. 
There exists degeneracy between $\Omega_{\rm}$ and the baryon fraction in the IGM $f_{\rm IGM}$. If observations, such as CMB and BBN, can provide a reliable measurement of $\Omega_{\rm b}$. Then we can constrain $f_{\rm IGM}$ using the ${\rm DM_{IGM}}-z$ relation \cite{Deng2014}. Li et al\cite{LiZX2019} propose a method of estimating $f_{\rm IGM}(z)$ using FRBs with DM and luminosity distance $d_{\rm L}$ measurements. Then they give a constrain of $f_{\rm IGM}=0.84^{+0.16}_{-0.22}$ with five localized FRBs \cite{LiZX2020}. 
Combining with other cosmological, such as CMB + baryon acoustic oscillations (BAO) + Supernova (SN) + $H_0$, probes can give strong constraints and it is effective for breaking the degeneracy between cosmological parameters and $f_{\rm IGM}$ \cite{Walters2019, Wei2019, WangB2023}.

\begin{figure*}
\centering
\includegraphics[width=\linewidth]{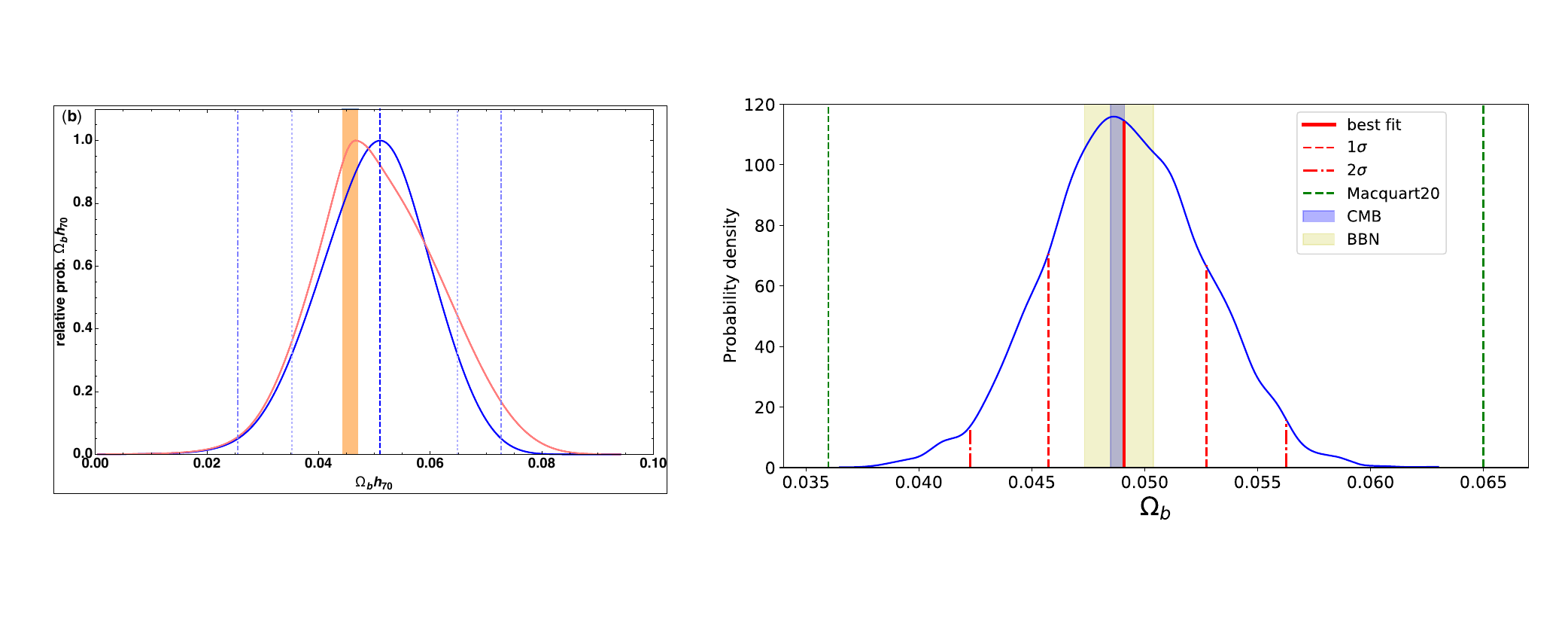}
\caption{{\bf Left panel: Density of cosmic baryons derived from five ASKAP-localized FRBs \cite{Macquart2020}. } 
The constraints on the IGM parameters $\Omega_b h_{70}$ and $F$, and the host galaxy parameters $\mu$ and $\sigma_{\rm host}$ for a log-normal DM distribution are derived using the five gold-standard bursts. The left panel displays the corner plot for $\Omega_b h_{70}$ where the orange shaded region denotes the range to which $\Omega_b h_{70}$ is confined by CMB+BBN measurements. The dotted and dot-dashed lines represent the 68\% and 95\% confidence intervals of each parameter respectively. The distribution of $\Omega_b$ is alternately marginalized over the range of $F$ indicated by cosmological simulations, $[0.09,0.32]$ (blue curve), and over the entire range of F $[0,0.5]$ investigated here (red curve). 
{\bf Right panel: The normalized probability density distribution of $\Omega_b$ from 22 localized FRBs \cite{Yang2022}.}
The best fit is $\Omega_b=0.0490^{+0.0036}_{-0.0033}$ in 1$\sigma$ confidence level. The red solid line shows the best-fit value, and dotted lines present the 1$\sigma$ and 2$\sigma$ confidence levels. The blue shade region shows the constraint on $\Omega_b$ from CMB with 1$\sigma$ confidence level, and the yellow shade region represents the constraint from BBN with 1$\sigma$ confidence level. The green dashed lines correspond to the 1$\sigma$ uncertainty range of $\Omega_b$ derived by Macquart et al \cite{Macquart2020}.
}
\label{fig_baryon}
\end{figure*}

\subsection{Measuring cosmological parameters}
\subsubsection{The equation of the state of dark energy}
\begin{figure*}
\centering
\includegraphics[width=1\linewidth]{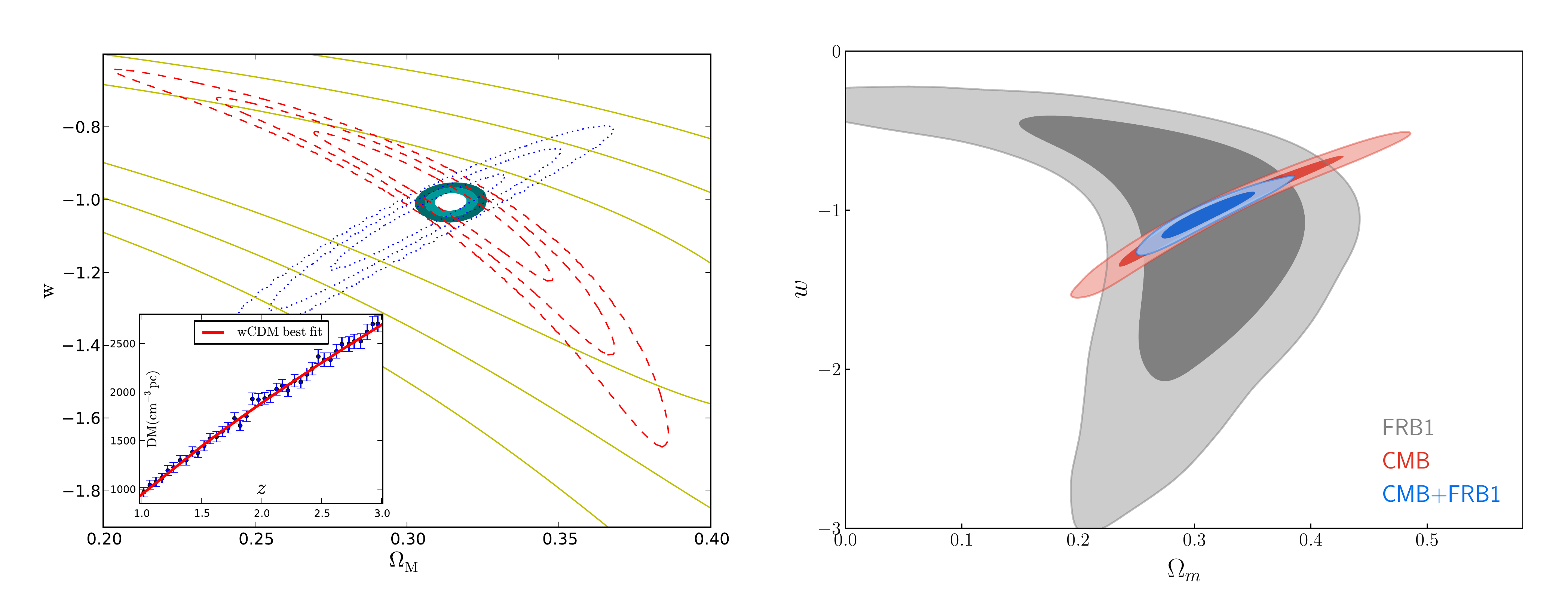}
\caption{{\bf Contours for the constraints of the state of dark energy $w$ and $\Omega_{m}$.} 
Left panel: The solid yellow lines, dotted blue lines, and dashed red lines are 580 SNe Ia, BAO data, and 1000 simulated FRBs. The shaded regions are the combined results \cite{ZhouB2014}. 
Right panel: Two-dimensional marginalized contours in the $\Omega_{m}-w$ plane for the $w$CDM model, using FRB, CMB, and CMB+FRB data \cite{Zhao2020}.
}
\label{fig_w}
\end{figure*}

The cosmological origin of FRB makes it a reliable cosmological probe, which provides an opportunity to measure the cosmological model parameters. One can give a constraint on the cosmological parameters in the ${\rm DM_{IGM}}-z$ relation using localized FRBs. 
Previous works explored the possibility of using FRBs as a viable cosmic probe through simulations \cite{ZhouB2014,GaoH2014,Walters2018}. 
The association between FRB and other astrophysical events has been supposed to constrain cosmological parameters. Gao et al\cite{GaoH2014} constructed a sample of FRB/GRB association systems to constrain the $w$CDM model combined with Type Ia supernova data. Gao et al\cite{GaoH2014} also proposed that the same method also applies to the FRB sample with z measured with other methods.
Wei et al\cite{Wei2018} simulated GW/FRB association systems and discussed the possibility of using it as a cosmological probe. 
Zhou et al\cite{ZhouB2014} discussed FRBs could help constrain the equation of the state of dark energy $w(z)$ (see left panel of Figure \ref{fig_w}). 
For high-redshift FRBs ($z>1$), the ${\rm DM_{IGM}}-z$ relation in a flat $w$CDM mode can be wrote as,  
\begin{equation}
    \langle{\rm DM_{\rm IGM}}(z)\rangle=\frac{3c\Omega_bH_0f_{\rm IGM}}{8\pi G m_p}\int_{0}^{z_{\rm FRB}}\frac{[Y_{\rm H}\chi_{\rm e,H}(z)+\frac{1}{2}Y_{\rm p}\chi_{e,He}(z)](1+z)}{\sqrt{\Omega_m(1+z)^3+ \Omega_{\rm DE}(1+z)^{3[1+w(z)]}}}dz, 
\end{equation}
where $\Omega_{\rm DE}$ is the density of dark energy in the universe. 
Kumar \& Linder \cite{Kumar2019} discussed the systematic uncertainties for accurate quantitative measurement.  

The joint measurement of FRB and other cosmological probes will improve measurement accuracy. Walters et al\cite{Walters2018} found the dark energy equation of state is poorly constrained using FRB + CMB + BAO + Supernovae (SNe) + $H_0$, but provides significant improvement for $\Omega_{b}h^2$. Zhao et al\cite{Zhao2020} proposed the combination of CMB and FRB data can improve the $w$ constraint using a large sample of simulated FRB data (see right panel of Figure \ref{fig_w}). 
For the upcoming Square Kilometre Array (SKA) era, several works propose that millions of FRBs with redshift measurements can tightly constrain the dark-energy equation of state parameters \cite{Qiu2022,ZhangJG2023}.

\subsubsection{Hubble constant}
Our universe is expanding with an expansion rate of $H(z)=\dot{a}(t)/a(t)$ ($a$ is the scale factor), which is the so-called Hubble parameter. In a flat $\Lambda$CDM universe, the Hubble parameter can be expressed as
\begin{equation}
    H(z) = H_0 \sqrt{\Omega_{\Lambda}+\Omega_{m}(1+z)^3}.
\end{equation}
$H_0$ is the expansion rate at the present day, which is called the Hubble constant. 
Wu et al\cite{Wu2020} first proposed a new method to measure the Hubble parameter $H(z)$ and they found that 500 mocked FRBs with dispersion measures and redshift information can accurately measure Hubble parameters using Monte Carlo simulation. 

\begin{figure}
\centering
\includegraphics[width=\linewidth]{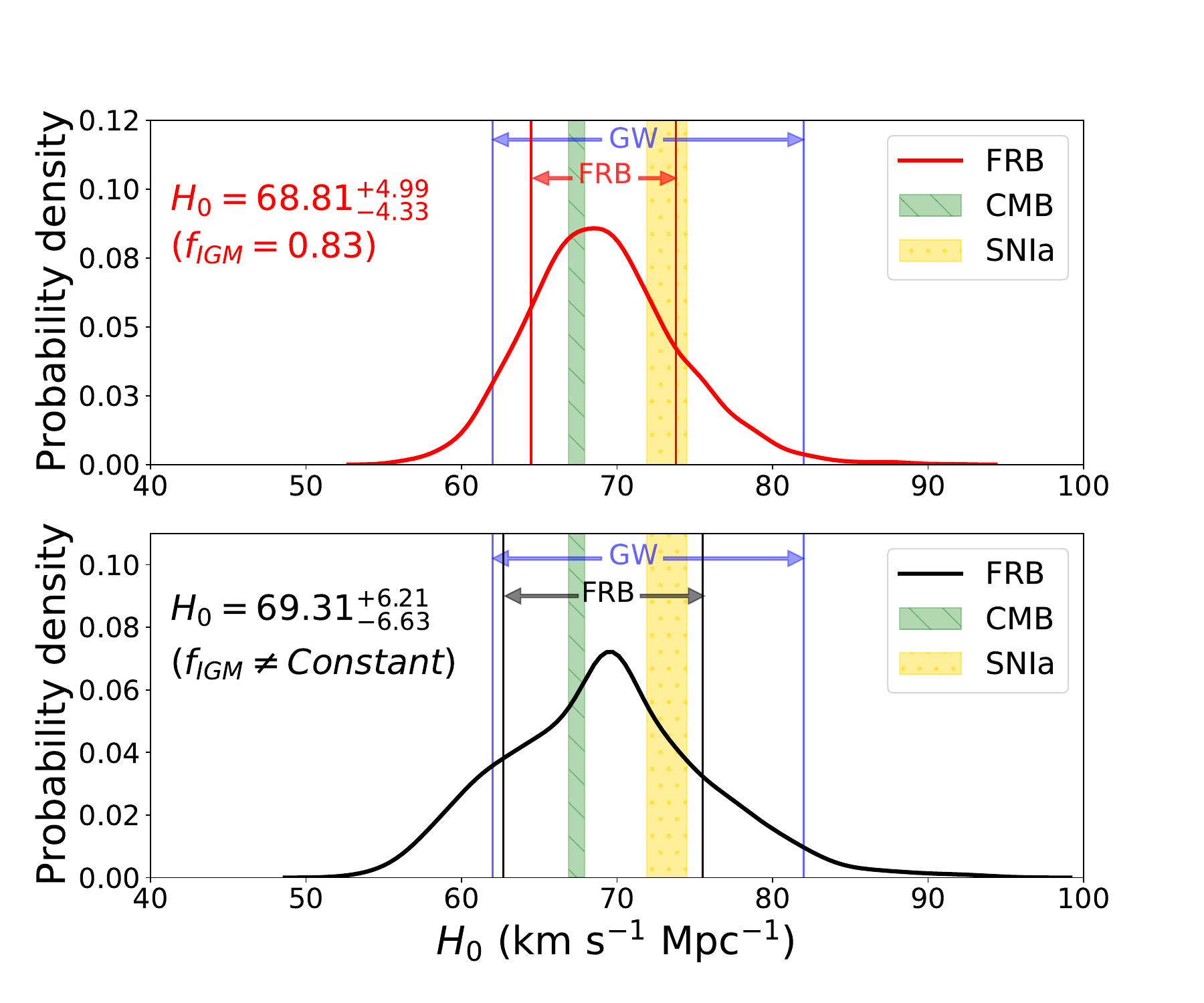}
\caption{{\bf The probability density distribution of $\bm H_0$ from eighteen localized FRBs \cite{Wu2022}.} The black solid line shows the probability density distribution of $H_0$. While the area between the two green vertical lines shows the result $H_0 = 62.86^{+5.01}_{-3.91} {\rm \ km \ s^{-1} \ Mpc^{-1}}$ with 1$\sigma$ uncertainty. The blue vertical lines correspond to the $H_0$ value derived by GW170817. The coral and yellow regions correspond to the 1$\sigma$ uncertainty range of $H_0$ reported by the Supernova H0 for the Equation of State (SH0ES) project and Planck, respectively.
}
\label{fig_H0}
\end{figure}

The Hubble constant is a basic parameter in cosmology, there are two mainstream measurement methods, the measurements of CMB by Planck Collaboration \cite{Planck2020} and the distance\-redshift relation of specific stars (e.g. Cepheid variables and type Ia supernovae) \cite{Riess2020, Riess2021}.
However, a significant difference is reflected in the Hubble constant measured by CMB and SNe Ia respectively, known as `Hubble tension' \cite{Freedman2017, Di Valentino2021}. 
FRB can provide an independent measurement method for measuring the Hubble constant. Hagstotz et al\cite{Hagstotz2022} gave a constrain of $H_0 = 62.3\pm 9.1\,{\rm km/s/Mpc}$ using nine FRBs with identified host counterpart and corresponding redshift. However, there exist systematic errors in $\rm DM_{host}$ and $\rm DM_{IGM}$ determination can not be ignored. 
Wu et al. \cite{Wu2022} first considered the $\rm DM_{host}$ and $\rm DM_{IGM}$ probability distributions from cosmological simulations \cite{Zhang2020,Zhang2021}. The reported a measurement of ${H_0} = 62.86^{+5.01}_{-3.91}{\rm\,km\,s^{-1}\,Mpc^{-1}}$ using eighteen localized FRBs, with an uncertainty of 7.9\% at 68.3 per cent confidence. Figure \ref{fig_H0} shows the measurement results considering the evolution of $f_{\rm IGM}$ with and without redshift, respectively. 
James et al\cite{James2022} gave a calculation of $H_0=73^{+13}_{-8}{\rm\,km\,s^{-1}\,Mpc^{-1}}$ utilize 16 ASKAP FRBs and 60 unlocalized FRBs from Parkes and ASKAP. 
Although there are only several FRBs identified in its host galaxy, simulations prospect that hundreds of FRBs will provide a high-precision measurement of $H_0$ without relying on other cosmological probes \cite{Hagstotz2022,Wu2022,James2022}. 
Identification of FRB host galaxy becomes a key observation strategy for FRB cosmology. Recent works have utilized more localized FRBs and combined them with other cosmological probes for $H_0$ measurement, further improving accuracy \cite{Wei2023,Liu2023,GaoJ2024}. 
Some works include unlocalized FRBs to measure constrain $H_0$ \cite{James2022, Zhao2022,GaoDH2024}, which offer an increase in both accuracy and precision beyond those obtained only by localized FRBs.

\subsection{Exploring the epoch of reionization history}

The epoch of reionization (EoR) marks a crucial phase in the history of the universe, during which the intergalactic medium, composed predominantly of neutral hydrogen (HI), was ionized by the earliest stars and galaxies. Exploring this epoch provides information on the very early universe and the first generation of stars and galaxies. Despite its significance in our comprehension of the universe, the timing, mechanisms, and structure of the EoR remain poorly defined. 
The dispersion measure in the IGM and redshift relation for radio afterglow of GRB firstly applies to measure the reionization history of the universe \cite{Ioka2003,Inoue2004}. 
High redshift FRBs are suitable candidates for exploring the reionization history \cite{Deng2014,Zheng2014}\cite{Linder2020,Bhattacharya2021,Fialkov2016,Caleb2019b,Beniamini2021,DaiJP2021,Hashimoto2021,Pagano2021,Zhang2021,Jing2022}.
Simulations predicted that a population of FRBs at $z\sim 3\text{-}4$ could serve as a tool to investigate the second helium reionization occurring approximately at redshift $z\approx 3.5$ \cite{Zheng2014, Linder2020, Bhattacharya2021}. While FRBs from $z>5$ could be used to probe the underlying astrophysics and morphology of the EoR and the optical depth of the Cosmic Microwave Background $\tau_{\rm CMB}$ \cite{Fialkov2016, Beniamini2021, DaiJP2021, Hashimoto2021, Pagano2021, Zhang2021}.

The dispersion measure from the intergalactic medium $\rm DM_{IGM}$ with fluctuations in electron density along the line of sight is the main component of $\rm DM_{obs}$. 
The mean $\rm DM_{IGM}$ corresponds to the reionization history and cosmological parameters,  
\begin{align}
\langle{\rm DM}_{\rm IGM}(z)\rangle =  \frac{3c}{8\pi G m_p} \int_0^{z} \frac{\Omega_b H_0 (1+z')  {\times[(1+f_{\rm He}) x_i(z) + f_{\rm He} x_i^{\rm HeII}(z)]}}{\sqrt{\Omega_m(1+z')^3+(1-\Omega_m)}} \,\mathrm{d}z',
\end{align}\label{eq:DMhomogcosmo}
where $H_0$ is the Hubble constant, $\Omega_m$ is the matter density parameter, $\Omega_b H_0^2$ is the physical baryon density parameter, $G$ is the gravitational constant, $m_p$ is the proton mass,and $f_{\rm He}$ is the number of helium relative to hydrogen atoms. $x_i^{\rm HeII}(z)$ is ionized fractions for the second helium reionization ($z=3\text{-}4$) and $x_i(z)$ quantifying the hydrogen (and first helium) reionization. 

\begin{figure}[t]
    \centering
    \includegraphics[width=\linewidth]{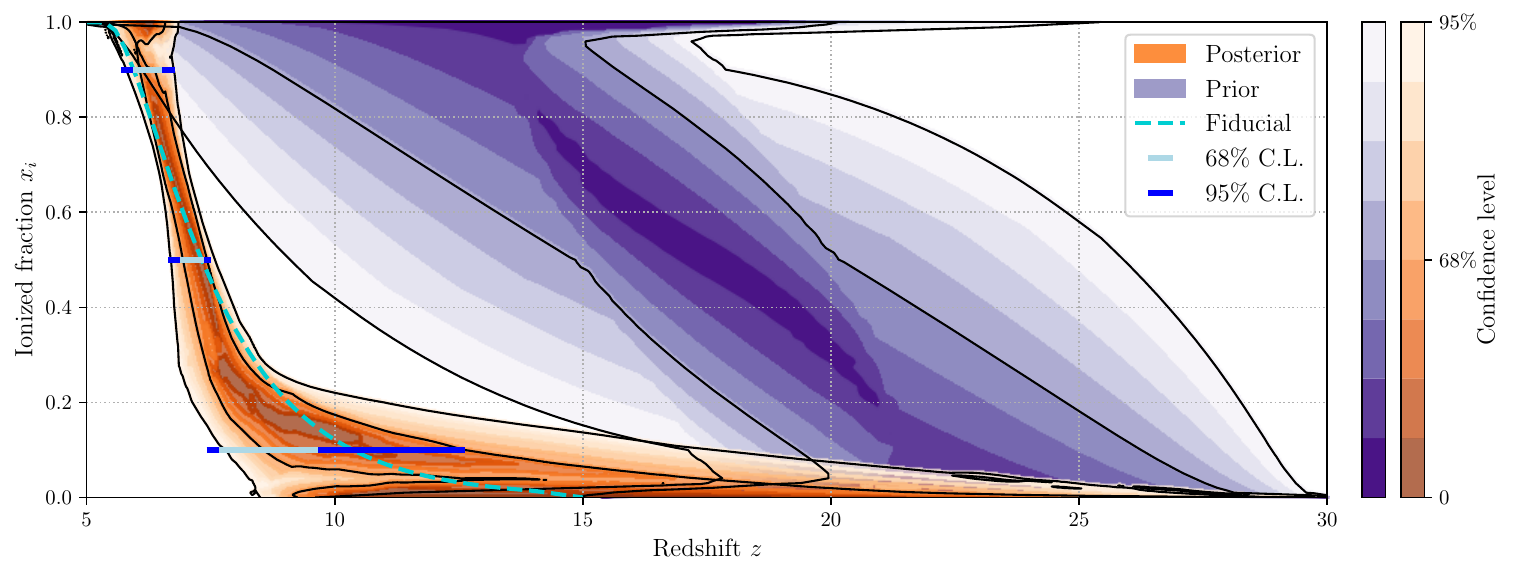}
    \caption{{\bf Posterior constraints on the total optical depth $\tau_{\rm CMB}$, marginalized over reionization histories as well as cosmological parameters $\Omega_b \cdot H_0$ and $\Omega_m$ \cite{Heimersheim2022}.}
    Here are the constraints from 100 and 1,000 FRBs, both including the $\tau_{15,30}$ limit.
    The measurement $\tau=0.0504^{+0.0050}_{-0.0079}$ (68\% confidence) is shown as black dashed line. FRBs can significantly improve the lower limit on the total optical depth constraint.} 
    \label{fig:reionization}
\end{figure}

For a sample high-redshift FRBs, the relation between the mean $\rm DM_{IGM}$ and $z$ depends on the full shape of reionization history $x_i(z)$. Heimersheim et al \cite{Heimersheim2022} simulated a sample of synthetic FRB, and they found that 1000 FRBs provide good constraints on the shape of the reionization history. The constraints on the ionized fraction $x_i(z)$ are demonstrated in Figure \ref{fig:reionization}, which gives full reionization history from redshift $z=5$ to $z=30$. The FRB data provides better constraints on the reionization history at $z<10$. 

The reionization of IGM poses challenges for measuring the CMB radiation propagating through the IGM. Planck Collaboration \cite{Planck2020} gives a result of the total optical depth to reionization $\tau_{\rm CMB} = 0.054 \pm 0.007$ (with 13\% uncertainty, 68\% confidence regions). 
The optical depth of the CMB is 
\begin{equation}
    \tau_{\rm CMB}(z) = \int_0^{z}\sigma_T n_e(z')dl,
\end{equation}
where $\sigma_T = 6.25\times 10^{-25}\,{\rm cm^2}$ is the Thompson cross-section. According to the definition of the dispersion measure,
\begin{equation}
    {\rm DM}(z) = \int_0^z\frac{n_e(z')}{1+z'}dl.
\end{equation}
Both $\tau_{\rm CMB}$ and $\rm{DM}(z)$ depends on the redshift evolution of electron density $n_e(z)$. And the relation between $\tau_{\rm CMB}$ and $\rm{DM}(z)$ 
\begin{equation}
    \tau(z)=\left[\frac{{\rm DM}(z)}{\rm cm^{-2}}(1+z)-\int_0^z \frac{{\rm DM}(z')}{\rm cm^{-2}}dz'  \right]\times \sigma_T.
\end{equation}
Knowing the redshift evolution of ${\rm DM}(z)$ will give an estimation of $\tau_{\rm CMB}$.
Heimersheim et al\cite{Heimersheim2022} simulated 100 and 1000 FRBs give constraints of optical depth with 11\% accuracy and 9\% accuracy, respectively. 
These methods provide a potential technique for exploring the reionization history. However, the lack of high-redshift FRBs challenges it. The upcoming Square Kilometre Array (SKA) era is expected to detect hundreds to thousands of FRBs at $z>6$ \cite{Fialkov2017,Hashimoto2020,Hashimoto2021,WeiJJ2024}, which will provide enough data in future observations.

\subsection{Probing the circumgalactic medium}

The circumgalactic medium (CGM) refers to the diffuse gas and plasma outside the disks and ISM of galaxies but within their virial radii, bounded by their gravitational forces \cite{Tumlinson2017}. In recent years, plenty of observational evidence has shown the important role CGM plays in galaxy evolution.
It can also be a possible solution to the well-known ``missing baryon" problem \cite{Fukugita2004}.
CGM is common and has been detected in wide ranges from high-redshifted quasars to local group galaxies. 
Despite its wide distribution, the study of CGM is quite challenging.
Frequent techniques for probing CGM include absorption-line studies, stacking, emission-line mapping, and hydrodynamic simulations.

As powerful, millisecond-duration bursts of radio emission from distant cosmic sources, FRBs will carry information about the CGM they propagate through \cite{Connor2022}.
The observed DM of FRB, $\rm DM_{obs}$, can be further subdivided into 
\begin{equation}
    {\rm{DM_{obs}}}(z) = {\rm DM_{MW,ISM}}+{\rm DM_{MW,halo}}+{\rm DM_{IGM}}(z)+\frac{{\rm DM_{host}}}{(1+z)}+\sum_{i}^{N_{\rm gal}}{\rm DM_{CGM,i}},
\end{equation}
where the CGM contributions from intervening galaxies ($\rm DM_{CGM,i}$) are summed over $N_{\rm gal}$ objects.
The CGM contributions may be included as part of the IGM cause baryons in the IGM are clustered in structures. Li et al\cite{LiY2019} have proposed that the CGM contributions from the foreground galaxies for nearby FRBs can be simply added.
The precise $\rm DM_{CGM,i}$ value for a given intervening galaxy hinges on the impact parameter of the background FRB ($b_{\perp}$) and the spatial distribution of gas density surrounding the galaxy, which remains unknown. 
While individual FRBs have been observed to traverse foreground halos \cite{Prochaska2019,Connor2020}, the lack of statistical data from a larger sample complicates the extraction of the CGM's DM contribution.

Thanks to the superb sensitivity of modern radio telescopes including the CHIME, large samples of FRBs are detected with wide coverages of redshifts and sky areas \cite{CHIME/FRB Collaboration2021}.
These FRBs provide vital information about the dispersion measure, rotation measure as well as spectra.
This information can provide constraints on key physical properties including the temperature, density, chemical composition, and the magnetic field.
Compared with traditional techniques such as the quasar absorption line, FRB will be more sensitive when dealing with ionized, magnetized, and low-redshift CGM. 
Also, point-like FRBs will be ideal sources due to their relatively clean, uniform background.

\begin{figure}
\centering
    \includegraphics[width=0.9\textwidth]{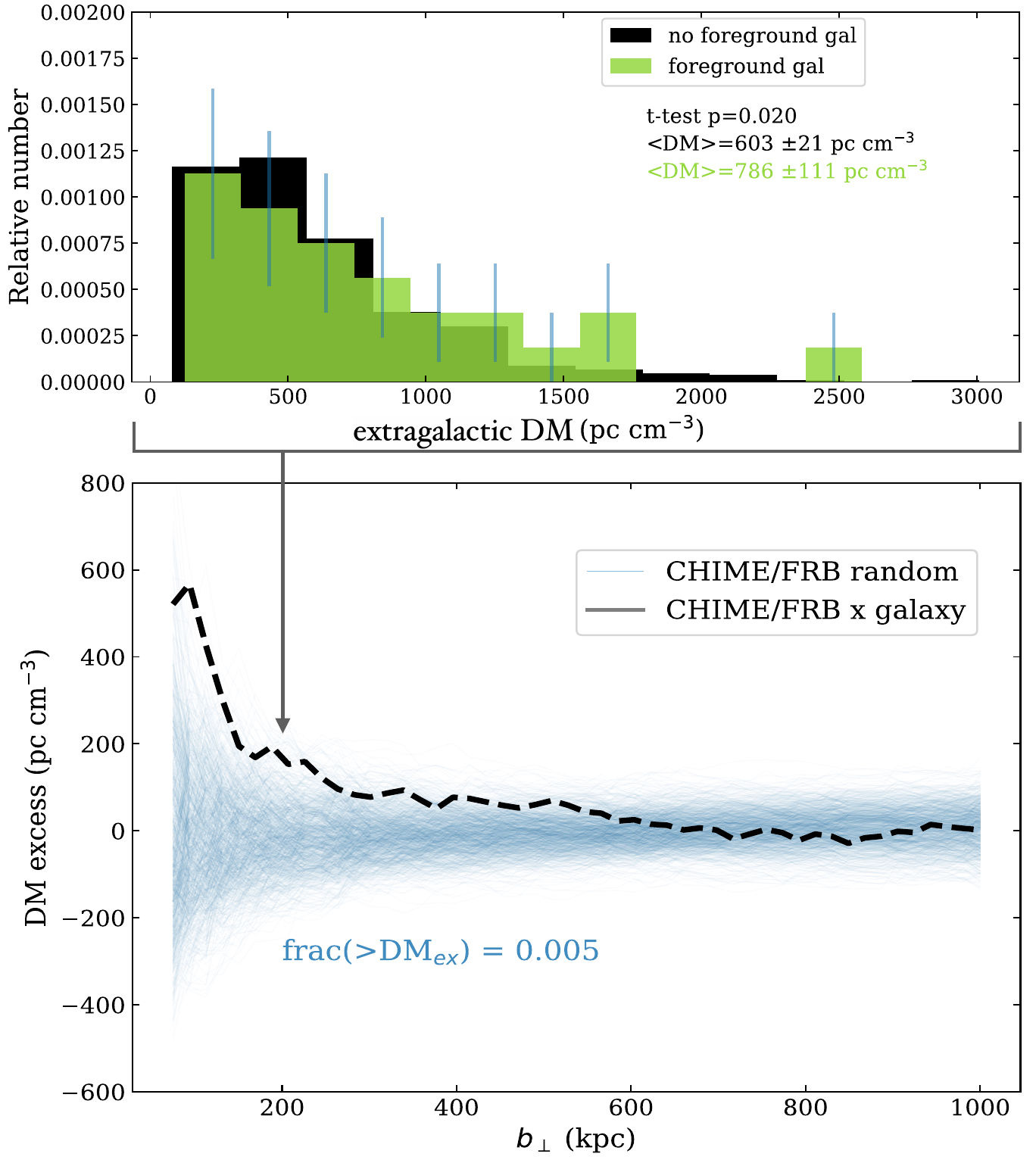}\\
    \caption{{\bf Statistical significance of the excess DM \cite{Connor2022}.} The top panel shows normalized extragalactic DM distributions of CHIME FRBs that intersect foreground galaxies (green) and those that do not (black) for a single fiducial value for the threshold impact parameter of $b_\perp=200$\,kpc, with $1\,\sigma$ Poissonian error bars. The vertical axis on the bottom panel shows excess DM as a function of the chosen threshold impact parameter $b_\perp$ for the full range. The excess is the difference in mean DMs between galaxy-intersecting and non-intersecting FRBs (black dashed curve). A jackknife test with a random shuffling of DMs among the CHIME/FRB sources (thin blue curves) is also shown for 1000 realizations. In just 0.5$\%$ of cases, the mean excess DMs of the random jackknife realizations are greater than the mean measured excess averaged between 75\,kpc and 300\,kpc.
    \label{fig:cgm}}
\end{figure}

A few efforts have been devoted to the statistical works based on CHIME/FRB Catalog 1 \cite{CHIME/FRB Collaboration2021}.
Based on 474 distant FRBs, Connor \& Ravi \cite{Connor2022} find that the mean DM of the galaxy-intersecting FRBs tend to have higher DM and the excess is $>~90{\rm~pc~ cm^{-3}}$ with $>95\%$ confidence. Figure \ref{fig:cgm} shows the excess DM's statistical significance.
It is claimed that although single FRBs can hardly be used to constrain their own CGM on their paths due to large uncertainties of different DM components, a large sample of FRBs will significantly aid in the modeling and interpretation of the CGM.
Despite these findings, several key technical challenges need to be addressed to effectively use FRBs to map the structure and evolution of the CGM across cosmic time.
First of all, the accurate localization of FRBs as well as the redshift measurements are crucial to place the CGM observations in a cosmological context.
The dispersion measure, rotation measure, and scintillation of FRBs contain contributions from other components including the Milky Way, the host galaxy, and the intergalactic medium.
Accurately modeling and subtracting these other contributions is necessary to isolate the CGM-specific signals.
So far, one still requires detailed theoretical models and simulations to establish the relation between the FRB-derived observables and the physical properties of the CGM.
More follow-up observations on more FRBs can also provide multi-wavelength information, thus leading to a more comprehensive understanding of these FRB-propagated CGM.

After all, the unique capabilities of FRB observations make them a valuable addition to the toolkit for studying the CGM. 
Wu \& McQuinn \cite{WuXH2023} measured DM excess for the CGMs of $10^{11}-10^{13}\,M_{\odot}$ halos using the CHIME/FRB first data release.
Connor et al\cite{Connor2023} proposed that the circumgalactic medium of intervening galaxies can be directly measured using FRB gravitational lensing. 
Medlock et al\cite{Medlock2024} probed the CGM with FRBs by analyzing the Cosmology and Astrophysics with Machine Learning Simulations using three suites, IllustrisTNG, SIMBA, and Astrid. 
As the field of FRB research continues to mature, the synergies between these different approaches will lead to a more comprehensive understanding of the role of CGM in galaxy formation and evolution.

\subsection{Gravitational lensing of FRBs}

The gravitational lensing of a fast radio burst will produce similar temporal patterns in observations. Lensed FRB systems provide an opportunity to probe the universe, such as probing dark matter \cite{Munoz2016,WangYK2018,Liao2020,Zhou2022,Chen2021,Ho2023,Connor2023,Katz2020}, measuring the cosmological parameters \cite{LiZX2018,Wucknitz2021,GaoR2022,Liu2019,Abadi2021}, testing weak equivalence principle \cite{YuH2018}, hunting axion dark matter \cite{GaoR2024,WangB2024}, and detecting cosmic strings \cite{Xiao2022}. 

\begin{figure}[t]
    \centering
    \includegraphics[width=\textwidth]{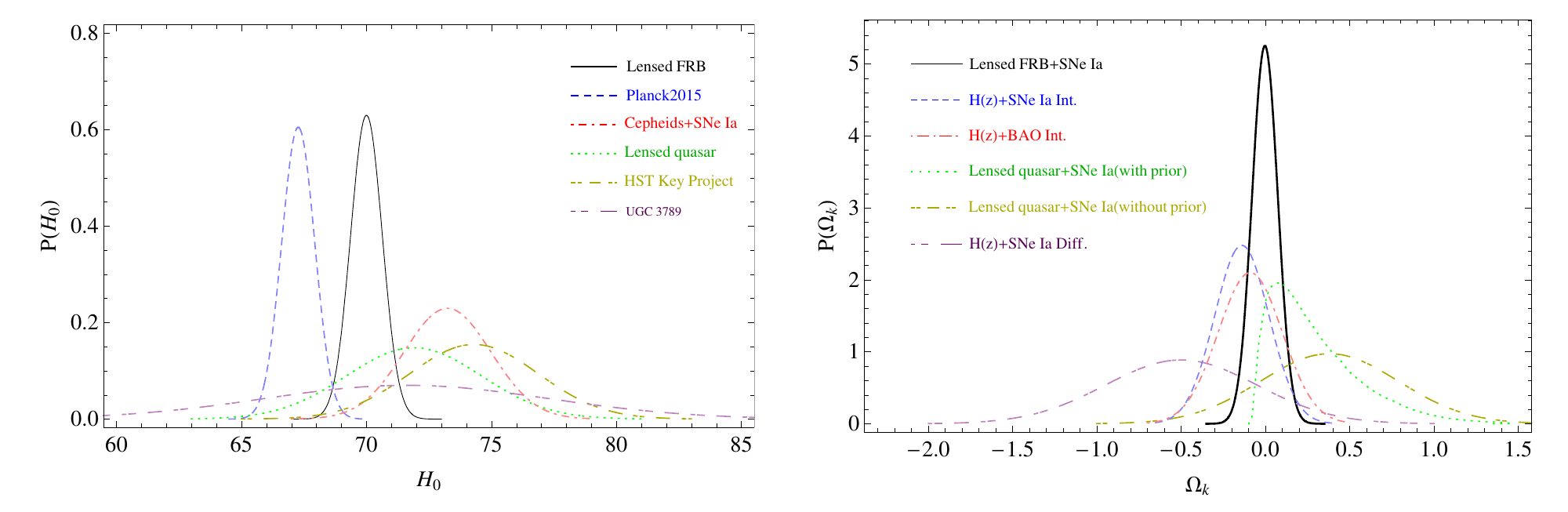}\\
    \caption{{\bf Probability distribution functions (PDFs) of the Hubble constant $H_0$ and the cosmic curvature $\Omega_k$ \cite{LiZX2018}.} Left panel: PDFs of the Hubble constant $H_0$ constrained from 10 lensed FRBs and some other currently available observations. Besides the result obtained from 10 lensed FRB systems (the black solid line), from top to bottom the lines represent $H_0$ inferred from the $Planck$ satellite CMB measurements,  local distance measurements, time-delay cosmography of strongly lensed quasars, distance measurements from the Hubble Space Telescope (HST) key project, and VLBI observations of water masers orbiting within the accretion disc of UCG 3789, respectively. 
    Right panel: Model-independent PDFs of the cosmic curvature estimated from 10 lensed FRBs and some other currently available observations. Besides the result obtained from 10 lensed FRB systems in this work (the black solid line), from top to bottom the lines are $\Omega_k$ inferred from the integral method with expansion rate (i.e., the Hubble parameter $H(z)$) and SNe Ia observations, the integral method with expansion rate and BAO observations, distance sum rule with the prior, distance sum rule without the prior $\Omega_k>-0.1$, and the differential approach with the expansion rate and SNe Ia observations, respectively.
 }\label{fig:lens}
\end{figure}

An FRB could be lensed by an isolated and extragalactic stellar-mass compact object \cite{Zheng2014,Munoz2016}, and an intervening galaxy \cite{LiZX2018,DaiL2017}. We will observe two images for a lensed FRB system. These two lensed images will exhibit slightly different characteristics in observations because they experience different paths. However, the time difference between the adjacent bursts will be the same for all lensed images \cite{LiZX2018}. 
According to the lensing theory, the time difference between two images $\Delta t$ is
\begin{equation}
    \Delta t = \frac{\Delta \Phi}{c}(1+z_l)\frac{D_l D_s}{D_{ls}},
\end{equation}\label{dt_lens}
where $\Delta\Phi$ is the Fermat potential difference between the two images, $c$ is the speed of light, $z_l$ is the redshift of the lens, $D_l$ is the angular diameter distance from the observer and to the lens, $D_s$ is the angular diameter distance from the observer and to the FRB source, and $D_{ls}$ is the angular diameter distance from the lens and to FRB source. 
The distance terms are inversely proportional to the Hubble constant $H_0$ and are dependent on the cosmic curvature $\Omega_k$. Li et al\cite{LiZX2018} proposed that 10 lensed FRB systems provide a precise estimation of $H_0$ and $\Omega_k$ in a model-independent way. Their constraints are shown in Figure \ref{fig:lens}. 

Mu{\~n}oz et al\cite{Munoz2016} proposed that an FRB lensed by a massive compact halo objects (MACHOs) lens with mass $M_L\sim 20-100\,M_{\bigodot}$ will generate two images with the milliseconds time difference. 
The next useful quantity for our purposes is the lensing optical depth, which is the probability that a source at redshift $z_s$ is lensed. 

The lensing optical depth of an FRB source, which is the probability that a source at redshift $z_s$ is lensed, can be expressed as 
\begin{equation}
    \tau(M_L,z_s) = \frac{3}{2}f_{\rm DM}\Omega_c\int_0^{z_s}dz_l\frac{H_0}{cH(z_s)}\frac{D_lD_{ls}}{D_s}(1+z_s)^2[\frac{(1+R_f)}{\sqrt{R_f}-2}-y_{\rm min}(M_L,z_s)],
\end{equation}
where $M_L$ is the mass of the lens, $\Omega_c$ is the cold-dark-matter density, $z_s$ is the redshift of the source, $f_{\rm DM}$ is the fraction of dark matter in MACHOs, $y_{\rm min}$ is the minimum normalized impact parameter, and $R_f$ is the flux ratio critical to ensure both events can be observed. The mass of the central lens is dependent on the optical depth. Given a distribution function for FRBs with a constant comoving number density and following the star-formation history, we can calculate the integrated optical depth. Then we can derive the number of lensed FRBs and the corresponding $f_{\rm DM}$ and $M_L$. Mu{\~n}oz et al\cite{Munoz2016} gave constraints $f_{\rm DM}\leq 0.08$ and $M_L\leq 20\,M_{\bigodot}$ using $10^4$ FRBs. 

However, there is still no observational evidence of FRB lensing systems, and it is not clear how many such systems will detect or how to find them. Recent works \cite{Connor2023} have forecasted detection rates for Deep Synoptic Array-2000 (DSA-2000), the Canadian Hydrogen Observatory and Radio transient Detector (CHORD), and the coherent all-sky monitor (CASM), as well as the CHIME/FRB survey, which is currently finding FRBs and searching for FRB lensing systems. They gave forecasts that 0.5-100 FRB lensing systems will be detected assuming 5\,yr operation with $\sim 80\%$ duty-cycle.

There are still other aspects of using FRBs as cosmological probes, such as constraining $f_{\rm IGM}$, IGM magnetic fields, turbulence and large-scale structure, weak equivalent principle WEP and rest photon mass, etc.
Other reviews \cite{Bhandari2021,ZhangB2023} provide detailed discussions on those effects.

\section{Challenges and future prospects}\label{sec:conclusion}

\subsection{Challenges}
There are still several unresolved questions in FRB physics, including the classification, progenitor engine, emission mechanism, trigger mechanism, propagation mechanism, and environmental factors of FRBs.

Currently, FRBs can be roughly classified into repeating and non-repeating FRBs based on whether they exhibit repeated bursts. However, those non-repeating FRBs may be misclassified as non-repeating due to observation selection effects and the sensitivity of telescopes. Therefore, long-term monitoring of clearly non-repeating FRBs can contribute to the classification of repeating and non-repeating bursts. 
Repeating FRBs being found to exhibit some unique observational characteristics, such as narrower spectra, complex time-frequency structure, longer duration, and down-drifting sub-bursts \cite{CHIME/FRB Collaboration2019b,Hessels2019,Pleunis2021b}. It remains uncertain whether these observed distinctions reflect an inherent difference between the two populations, a propagation effect influenced by distinct source environments, or a selection bias. There still are possibilities that all FRBs repeat and the repetition intervals span a wide range, or non-repeating FRBs are a distinct population different from known repeaters, or there are other classes of FRBs \cite{Ai2021,Lin2023b}. 

The discovery of FRB 20200428 originating from a magnetar SGR J1935+2154 within the Milky Way suggests that at least some cosmological origins could also be magnetars. Other possible models cannot be completely ruled out, indicating that FRBs may still have multiple origins. For non-repeating FRBs, some studies suggest they have cataclysmic progenitor models \cite{Lieu2017,Barrau2014,Barrau2018,Platts2019,ZhangB2023}. 
The discussion on the emission site in the magnetar model is controversial, with the main viewpoints divided into inside \cite{Katz2014,Katz2018,Katz2020b,Kumar2017,Lu2018,Yang2018,Yang2020,Lu2020,WangJS2020,Cooper2021,WangWY2022a,WangWY2022b,Zhang2022,Qu2023} and outside the magnetosphere \cite{Lyubarsky2014,Metzger2017, Beloborodov2017,Beloborodov2020}. 
The extremely high brightness temperatures of FRBs require that their radiation mechanism is coherent emission. The radiation mechanisms widely discussed in the field of FRBs include coherent curvature radiation mechanism \cite{Katz2014,Katz2018,Katz2020b,Kumar2017,Lu2018,Yang2018,Yang2020,Lu2020,WangJS2020,Cooper2021,WangWY2022a,WangWY2022b}, coherent inverse Compton scattering emission \cite{Zhang2022,Qu2023,Qu2024}, relativistic shock emission \cite{Lyubarsky2014,Beloborodov2017,Beloborodov2020,WuQ2020,Metzger2019,Margalit2020a,Margalit2020b,YuYW2021}, and coherent Cherenkov radiation\cite{LiuZN2023}.
The above models can explain the millisecond duration, peak frequency, and high luminosity of FRBs. However, there are some polarization properties that are difficult to explain in some models.
Ninety percent of circular polarization has been detected in FRB 20201124A \cite{JiangJC2024}. Zhao et al\cite{ZhaoZY2024} proposed that high circular polarization arises from magnetospheric propagation effects caused by relativistic plasma. 
Niu et al\cite{NiuJR2024} found some sudden polarization angle jumps of FRB 20201124A, which challenges some far-away synchrotron maser shock models.

Another confusing aspect is the local environment of FRBs. Repeating FRB 20121102A, FRB 20190520B, and FRB 20201124A associated with a persistent radio source \cite{Chatterjee2017,NiuCH2022,Bruni2024}, indicate the presence of dense nebulae near FRB origins. 
Recently, Ibik et al \cite{Ibik2024} have identified two candidate PRSs (FRB 20181030A and FRB 20190417A) from the 37 CHIME/FRB repeaters. 
The relation between the persistent emission luminosity and the rotation measure has been proposed \cite{YangYP2020,Bruni2024,Ibik2024}. 
This has prompted speculation about associations with supernova remnants or pulsar/magnetar wind nebulae in at least some FRB sources \cite{LiQC2020}. 
The observed periodicity in FRB 20180916B \cite{CHIME/FRB Collaboration2020a}, and possibly FRB 20121102A \cite{Rajwade2020}, has sparked speculation about a binary environment for some FRB sources \cite{Ioka2020}. Furthermore, rapid variations in RMs \cite{FengY2022} in active repeaters such as FRB 20201124A \cite{Xu2022} and FRB 20190520B \cite{Anna-Thomas2023} suggest a dynamically evolving magnetized environment surrounding these FRBs \cite{WangFY2022}. 

The research on these issues is expected to persist for years, much like developments in other fields of astronomy, with the potential for these issues to be resolved in the future.

\subsection{Prospects}

Since the discovery of FRBs in 2007, less than 20 years ago, the field of FRB has rapidly advanced compared to the fields of pulsars and gamma-ray bursts. Observations and theoretical research on FRBs are undergoing a phase of rapid development.

1. The number of observed FRB sources is expected to grow rapidly. Zhang\cite{ZhangB2023} and Petroff et al\cite{Petroff2022} predicted that within the next few years of observations, the total number of FRB sources will reach ~10,000. Such optimistic expectations are based on projects like the CHIME/FRB project. The Square Kilometre Array (SKA) is also expected to discover more FRBs in the future, with estimates ranging from $10^4$ to $10^6$ detections \cite{Fialkov2017}. A large sample is crucial for the statistical analysis of FRBs, including the population synthesis studies and classifications of FRBs. 

2. Long-term monitoring of active repeating FRBs will accumulate more repeating bursts. Radio telescopes with high sensitivity, such as the FAST, will be capable of detecting more of these repeating FRBs. For the identified CHIME/FRB, ASKAP, Parkes, DSA and FAST-discovered repeating FRBs, long-term monitoring of these repeaters will trace the possible dynamic evolution of the properties of FRBs or their local environments, such as DM and RM reflecting the propagation processes \cite{Xu2022,Anna-Thomas2023}. It is crucial to understand the origin of FRBs and their surrounding environments. In the next few years, long-term monitoring of RM and DM may show quasi-periodic behavior, which is robust evidence for the binary origin of FRBs \cite{WangFY2022,Zhao2023}. 

3. Localization of FRBs has always been a critically important topic in the field of FRBs. With observations from more sensitive and wider-field radio telescopes, an increasing number of FRB host galaxies are being confirmed. However, compared to the total number of FRB sources, localized FRBs represent only a small fraction. Current observational facilities such as ASKAP-CRAFT system \cite{Aggarwal2021b}, MeerKAT telescope \cite{Sanidas2018}, the VLA Low-band Ionosphere and Transient (VLITE)-Fast experiment \cite{Ray2021}, the Apertif LOFAR Exploration of the Radio Transient Sky (ALERT) program \cite{Maan2017}, and UTMOST-2D project of the Molonglo Radio telescope \cite{Deller2020} are capable of further improving the localization precision of FRBs while gradually increasing the number of localized FRBs. In the future, with the advent of the DSA-2000 \cite{Hallinan2019}, the Hydrogen Intensity and Real-time Analysis eXperiment (HIRAX) \cite{Newburgh2016}, CHORD \cite{Vanderlinde2019} and the SKA eras \cite{Fialkov2017}, the number of localized FRBs may reach between $10^3$ and $10^4$, providing advancements in the statistical study of host galaxies and cosmological applications.  

4. Multi-wavelength observations and multi-messenger counterparts of FRBs are crucial for understanding their origins and radiation mechanisms. The association of FRB 20200428 with an X-ray burst within the Milky Way \cite{CHIME/FRB Collaboration2020b,Bochenek2020,LiCK2021,Mereghetti2020,Ridnaia2021,Tavani2021} helped constrain the radiation mechanism of the magnetar-origin FRBs. However, so far, no extragalactic FRB has been discovered with multi-messenger events. The discovery of such associated events would greatly advance observational and theoretical research.

\section*{Acknowledgements}
This work thanks the two reviewers for their valuable suggestions.
This work was supported by the Postdoctoral Fellowship Program of CPSF under Grant Number GZB20240308 and the National Natural Science Foundation of China (grant No. 12273009).  

\end{document}